\documentclass[journal, twocolumn]{IEEEtran}
\usepackage{graphicx} 
\usepackage{tabularx}
\usepackage{hyperref}
\usepackage{caption}
\usepackage{subcaption}
\usepackage{amsmath}  
\usepackage{amssymb}  
\usepackage{booktabs}
\usepackage{tabularx}
\usepackage{verbatim}
\usepackage{multirow}
\usepackage{longtable}
\newcolumntype{M}[1]{>{\centering\arraybackslash}m{#1}}
\usepackage{float}
\usepackage{color, soul}
\sethlcolor{yellow}
\usepackage{algorithmic}
\usepackage{xcolor}
\usepackage{algorithm}
\newcolumntype{M}[1]{>{\centering\arraybackslash}m{#1}}
\usepackage{tabularx}
\usepackage{enumitem}
\newcolumntype{C}[1]{>{\centering\arraybackslash}m{#1}}

\setlist[itemize]{leftmargin=*}

\setlist{leftmargin=*}

\begin{document}

\title{A Tutorial-cum-Survey on Self-Supervised Learning for Wi-Fi Sensing:  Trends, Challenges, and Outlook}
\author{Ahmed Y. Radwan, Mustafa Yildirim, Navid Hasanzadeh, Hina Tabassum, {\em Senior Member IEEE}, and Shahrokh Valaee, {\em Fellow IEEE}\thanks{ A. Radwan, M. Yildirim, and H.~Tabassum are with the Department of Electrical Engineering and Computer Science, York University, ON, Canada. (E-mails: ahmedyra@yorku.ca, yildrim@my.yorku.ca, hinat@yorku.ca).

\par H.~Tabassum is a visiting faculty  with the Edward S. Rogers Sr. Department of Electrical and Computer Engineering, University of Toronto, Canada. 

\par N. Hasanzadeh and S. Valaee are with the Edward S. Rogers Sr. Department of Electrical and Computer Engineering, University of Toronto, Canada.
(E-mails: navid.hasanzadeh@mail.utoronto.ca,  valaee@ece.utoronto.ca)
}}

\maketitle
\begin{abstract}
Wi-Fi technology has evolved from simple communication routers to sensing devices. Wi-Fi sensing leverages conventional Wi-Fi transmissions to extract and analyze channel state information (CSI) for applications like proximity detection, occupancy detection, activity recognition, and health monitoring. By leveraging existing infrastructure, Wi-Fi sensing offers a privacy-preserving, non-intrusive, and cost-effective solution which, unlike cameras, is not sensitive to lighting conditions.
Beginning with a comprehensive review of the Wi-Fi standardization activities,  this tutorial-cum-survey  first introduces fundamental concepts related to Wi-Fi CSI, outlines the CSI measurement methods, and examines the impact of mobile objects on CSI.
The mechanics of a simplified testbed  for CSI extraction are also described. Then, we present a qualitative comparison of the existing Wi-Fi sensing datasets, their specifications, and pin-point their shortcomings. Next, a variety of preprocessing techniques are discussed that are beneficial for feature extraction and explainability of  machine learning (ML) algorithms. We then provide a qualitative review of recent ML approaches in the domain of Wi-Fi sensing and present the significance of self-supervised learning (SSL) in that context. Specifically, the mechanics of contrastive and non-contrastive learning solutions is elaborated in detail and a quantitative comparative analysis is presented in terms of classification accuracy. {Finally, the article concludes by highlighting  emerging technologies that can be leveraged to enhance the performance of Wi-Fi sensing and  opportunities for further research in this domain.}
\end{abstract}


\raggedbottom

\section{Introduction}

In recent years, the rapid expansion of Wi-Fi infrastructure has driven advancements in its applications beyond simple connectivity. As Wi-Fi or radio frequency (RF) signals traverse physical spaces, they interact with objects and people, causing reflection, diffraction, and scattering. These interactions generate multi-path effects, revealing details about the environment, such as human movements, positions, and object conditions \cite{9900419}. These details are typically decoded using two key metrics: \textit{Channel State Information (CSI)} and \textit{Received Signal Strength Indicator (RSSI)}. CSI is a complex-valued matrix representing the Channel Frequency Response (CFR) influenced by factors like distance, multi-path effects, and the Doppler effect. RSSI measures the power level of the received signal, which is influenced by  distance and obstacles. In addition,
\textit{CSI similarity}, calculated through cross-correlation of the two CSI matrices, is also a popular metric in motion-related Wi-Fi sensing applications, effectively distinguishing between static and moving objects. Furthermore, \textit{channel coherence time} and \textit{coherence bandwidth}, representing the duration or bandwidth during which the Channel Impulse Response (CIR) remains stable, are also used to detect the mobility of Wi-Fi devices.

Key applications include crowd counting, occupancy detection, pose estimation, vital signs monitoring,  human activity, and gesture recognition \cite{10.1145/3310194}. A comprehensive summary of Wi-Fi sensing applications is presented in \textbf{Table~1}.

\begin{figure*}[h]
	\centering
	\subfloat{%
		\includegraphics[width=1.0\linewidth]{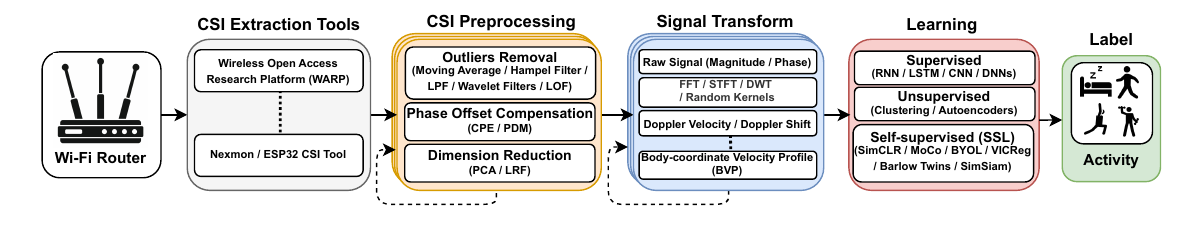}
	}
	\caption{HAR using Wi-Fi CSI}
\label{figure:diagram}
\end{figure*}

\begin{table*}
\centering
\caption{Wi-Fi Sensing Applications}
\label{applications_table}
\resizebox{\textwidth}{!}{
\small
\begin{tabularx}{\textwidth}{@{} l l X @{}}
\toprule
\textbf{Application} & \textbf{References} & \textbf{Description} \\
\midrule
Localization & \cite{li2021deep}, \cite{8397121}, \cite{zhou2020adaptive}, \cite{yang2021indoor}, \cite{zhang2021low} & 
Tracking a target's location within a space using ambient Wi-Fi signals. Traditional methods require the target to have transmitting or receiving hardware. Recent studies focus on device-free tracking using CSI data to create environmental signal fingerprints. Techniques like Domain Adaptation and Transfer Learning are used to address accuracy issues due to environmental changes. \\
\midrule
Human Activity Recognition (HAR) & \cite{9900419}, \cite{shi2022environment} & 
Recognizes activities such as sitting, standing, walking, and running. Useful for monitoring room occupancy in smart homes. Fall detection safeguards elderly or ill individuals without invading privacy, unlike camera-based systems. \\
\midrule
Gesture Recognition & \cite{10.1145/2942358.2942393}, \cite{9233449}, \cite{8416308}, \cite{10.1145/3191755} & 
Identifies intricate human movements focusing on hand and finger gestures. Enables innovative gesture-based interactions in smart homes and in-vehicle controls. Sign Language detection is another use case. \\
\midrule
Crowd Counting and Occupancy Detection & \cite{9900419}, \cite{di2016trained} & 
Estimates the number of people in a given area. Detects stationary crowd sizes by subtle, random fidgeting movements. Valuable for safety-critical and non-critical situations. Used to monitor human queues or tally individuals in waiting rooms. Helps improve services in various locations like retail stores, airports, hospitals, and theme parks. Optimizes transportation schedules and boarding/payment procedures based on passenger numbers. \\
\midrule
Health Monitoring & \cite{zeng2019farsense}, \cite{9831898}, \cite{10.1145/2746285.2746303}, \cite{wang2017phasebeat}, \cite{shirakami2021heart} & 
Continuous health monitoring in private residences. Common applications include respiration tracking and detecting irregular breathing patterns, such as apnea or tachypnea, especially in sleep. Tracks heart rates to reveal heart rhythm variability. \\
\bottomrule
\end{tabularx}
}
\end{table*}

\begin{figure*}[h]
	\centering
	\subfloat{%
		\includegraphics[width=1.0\linewidth]{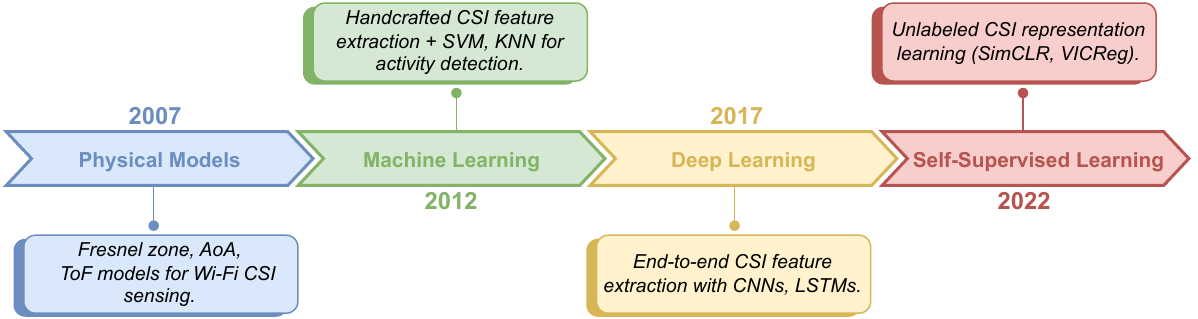}
	}
	\caption{Evolution of Methodologies for Wi-Fi Sensing}
\label{figure:evolution_wifisensing}
\end{figure*}
Wi-Fi sensing provides an effective solution for indoor sensing and tracking, offering several advantages over traditional methods such as motion sensors, infrared systems, wearable devices, and camera-based systems. It leverages the widespread availability of Wi-Fi infrastructure, avoiding the need for additional hardware. Unlike wearable sensors, Wi-Fi sensing is non-intrusive, as it does not require individuals to wear or carry any devices. It also enhances privacy compared to cameras, as the CSI data collected does not contain facial or identifiable visual information, thus addressing privacy concerns effectively \cite{Yang_Chen_Wang_Zou_Lu_Sun_Xie_2022}.
Additionally, Wi-Fi sensing can operate in non-line-of-sight (NLOS) settings, giving it an advantage over camera-based systems that require a direct line of sight (LOS). This ability to sense through walls and obstacles makes Wi-Fi sensing more versatile for various indoor applications. Furthermore, it is more cost-effective and computationally efficient than camera-based solutions. Cameras often need multiple units to cover different areas and are prone to lighting conditions, whereas infrared systems are prone to high cost and processing demands. Wi-Fi sensing, on the other hand, can make use of existing routers, reducing these limitations and providing a more streamlined solution \cite{9729463}.




{The end-to-end process of Wi-Fi sensing is shown in \textbf{Fig.~\ref{figure:diagram}}. Beginning with the Wi-Fi routers, the selected hardware enables CSI extraction, which due to its inherently noisy nature, undergoes preprocessing techniques. Following preprocessing, signal transformation methods are typically applied to extract meaningful features from raw CSI measurements. Finally, the data is used by Wi-Fi sensing algorithms to accomplish the downstream tasks, such as human activity recognition (HAR).}

To date, Wi-Fi sensing algorithms can be broadly categorized into two primary groups: \textit{model-based} and \textit{ML-based} methods. {The historical evolution of these methodologies is summarized in\textbf{ Fig.~\ref{figure:evolution_wifisensing}}, starting from early model-based methods to modern deep learning and self-supervised learning approaches.} Model-based methods utilize physical models, such as the Fresnel Zone, Angle of Arrival (AoA), and Time of Flight (ToF), to characterize amplitude attenuation and phase shifts in multipath channels based on direction and distance. Common techniques in this category include the Multiple Signal Classification (MUSIC) algorithm, which estimates AoA, and the Power Delay Profile (PDP), obtained through the Inverse Fast Fourier Transform (IFFT) of CSI, for ToF estimation. These methods offer valuable insights into the underlying mechanisms of Wi-Fi sensing, facilitating applications like respiration monitoring and fall detection.

However, the \textit{model-based methods} heavily depend on statistics to represent human activities and signal variations, which becomes challenging when handling complex movements. Particularly, for HAR, model-based methods often struggle to accurately model the correlations between intricate actions and corresponding changes in CSI. Furthermore, these methods face performance degradation in NLOS environments where reflections and obstructions complicate signal interpretation. 
{Moreover, \textit{model-based methods} assume specific stationary data distributions for each class or label. While this simplification makes the model less complex and more interpretable, Wi-Fi signals and the effects of environmental noise typically do not follow well-known statistical distributions in practice. In addition, model-based methods often treat each CSI sample independently, without considering temporal dependencies. Thus, these model characteristics hinder their generalization across different settings, necessitating specific calibration for each new environment.}

Additional examples of model-based algorithms include threshold-based methods and clustering. These approaches often incorporate statistical metrics such as variance, Mean Absolute Deviation (MAD), and Power Spectral Density (PSD). Threshold-based algorithms rely on predefined thresholds for activity detection. However, a threshold suitable for one activity often proves unsuitable for others.
Thus, when multiple activities occur within a single detection window, it is challenging to select a universal threshold without prior knowledge of activity granularity \cite{chen2022lightseg}.


Unlike model-based methods that rely on predefined physical models and statistical assumptions, machine learning (ML) based algorithms can infer complex patterns in data directly from observed CSI signals. This capability makes them adept at handling intricate human movements and adapting to diverse environments, including challenging NLOS scenarios. The flexibility of ML-based methods allows for modeling temporal dependencies and uncovering latent representations that may not conform to traditional statistical distributions. These advantages enable ML-based algorithms to generalize more effectively across different conditions and users, reducing the need for environment-specific calibration and making them highly suitable for dynamic real-world applications. 

{However, ML-based methods typically require large amounts of labeled data, which is challenging. This dependency makes training ML models, especially supervised ones, difficult and underscores the need for solutions that effectively utilize limited amount of labeled data.} Therefore, the \textbf{scope of this tutorial-cum-survey article} is to perform a comprehensive review of the existing research works in the domain of machine-learning empowered Wi-Fi sensing, with a focus on unsupervised and self-supervised learning  (SSL) techniques in which the requirement of labeled data is minimal.

\subsection{Existing Survey/Tutorial Articles}
In this subsection, we classified and discussed the existing survey articles into two classes, i.e.,\textit{ model-based surveys and ML-based surveys}
\subsubsection{Model-Based Surveys}
One of the earliest studies in this area \cite{7040509} laid the foundation by demonstrating the potential of Wi-Fi sensing. This survey provided a detailed comparison of CSI and RSSI  and their utilization for different tasks, at a time when Wi-Fi sensing was still in its infancy and model-based approaches were yet to be explored. 
Later, \cite{10.1093/jcde/qwab003} offered a comprehensive survey focused on model-based techniques, marking it as one of the earliest works in this direction. They provided a flow analysis for CSI extracting and demonstrated how CSI could be used for subsequent tasks. Additionally,  various methods, such as CSI-Mobility, CSI-Ratio, and CSI-Quotient were discussed.

\subsubsection{ML-Based Surveys}
{A broader survey on machine learning-enhanced Integrated Sensing and Communication (ISAC) systems was presented in \cite{ISACrespati2024survey}. This work systematically reviewed various ISAC system configurations, sensing sources, and real-world applications, while emphasizing how ML and DL techniques can enhance both sensing tasks (e.g., localization, gesture recognition) and optimize system operations (e.g., channel estimation, waveform design, and beamforming). Unlike surveys focused specifically on Wi-Fi sensing, it provided a general framework situating ML across a wider range of ISAC applications for 6G networks.}
In \cite{9900419}, key components of edge-based Wi-Fi sensing were introduced, including in-depth signal processing, data preparation, and prediction-making. It also highlighted the use of a tree-structured Parzen Estimator for hyperparameter optimization and demonstrated the capability of ESP32 microcontrollers to process CSI data and perform on-device inference efficiently.
In \cite{10.1145/3310194}, the authors presented the first survey that systematically categorized Wi-Fi sensing techniques into detection, recognition, and estimation applications, while providing a unified analysis framework for signal processing techniques and algorithmic approaches. Their key contribution was establishing a systematic comparison of modeling-based, learning-based, and hybrid approaches, along with identifying critical research challenges in robustness, privacy, and the coexistence of sensing with networking.

Another survey \cite{9076313} highlighted the benefits of deep learning (DL), such as using convolutional neural networks (CNNs) for spatial features and long-short term memory (LSTMs) for temporal features. This survey reviewed various ML-based approaches and preprocessing techniques to address the drawbacks of model-based methods. Additionally, \cite{9729463} provided a comprehensive analysis of Wi-Fi sensing in healthcare settings, focusing on detecting critical health events like falls, sleep disturbances, and abnormal vital signs. They also discussed the challenges of diverse data, privacy concerns, and system requirements, offering insights for future improvements in healthcare applications.
Furthermore, \cite{ahmadiftikhar2024wifi} provided a review of DL applications in Wi-Fi sensing. They highlighted setups used in recent papers for various activities, emphasizing DL's capability to improve accuracy and efficiency in human-based tasks. Specifically, the survey evaluated DL architectures, including hybrid models like transformers and autoencoders, providing insights into the effectiveness of combining these architectures for enhanced performance.

Recently, in \cite{xu2023selfsupervisedlearningwificsibased}, the authors presented a  systematic study of SSL in Wi-Fi CSI-based HAR. Their work provided insights into SSL framework implementation, from pre-training to fine-tuning, and evaluated different SSL architectures  on the Widar~\cite{9516988} dataset. {Our survey complements this work by providing broader coverage of the Wi-Fi sensing pipeline, including detailed analysis of data collection methodologies, hardware setups, and preprocessing techniques}. While their study focused on evaluating SSL robustness across different environments within the same dataset, {our work extends the analysis to cross-dataset and cross-task scenarios, exploring domain adaptation between different sensing applications}. {We also examine both contrastive and non-contrastive SSL methods. Additionally, our survey provides implementation guidelines for various deployment scenarios and investigates SSL performance across a diverse range of publicly available Wi-Fi sensing datasets}. 

{While also in \cite{xu2025evaluating} they offer a comprehensive evaluation of SSL, their main focus is solely on SSL, unlike our broader scope that includes few-shot learning on top of SSL for downstream tasks.} {Moreover, our cross-domain evaluations are designed for more extreme scenarios to assess the generalizability of SSL—testing across different environments, tasks, and datasets—whereas their domain shift evaluation is limited to intra-dataset settings such as user and room changes within Widar and CSIDA~\cite{CSIDA}.} {This distinction highlights our contribution toward understanding SSL deployment under realistic and challenging domain shifts.}

{A comparative analysis of the existing survey articles in terms of {their primary focus on modeling and methodology, employed signal-processing and DL techniques, and their key contributions and limitations} is provided in }\textbf{Table~\ref{rel_works1}}.

\subsection{Motivation and Contributions}

Existing surveys in the domain have extensively explored the potential of Wi-Fi-based applications and highlighted their advantages over traditional methods. Recognizing the significant promise that Wi-Fi sensing offers, this survey-cum-tutorial shifts the focus towards emphasizing the significance of ML techniques and, in particular, SSL for Wi-Fi sensing. This focus is motivated by the need to address the challenges posed by \textbf{limited availability of labeled data} and the \textbf{lack of generalization} across environments.
The key contributions of this article can be summarized as follows:
\begin{itemize}
   \item {We present a comprehensive review of end-to-end Wi-Fi sensing networks, spanning from data collection {and CSI processing} methods to enhance DL solutions, with an emphasis on SSL applications for few-shot learning and cross-domain adaptation.}

   \item {We present an in-depth analysis of Wi-Fi CSI behavior for moving users, including the mathematical modeling of motion-induced effects, Doppler velocity estimation, and a review of unsupervised high-dimensional representation learning approaches for feature extraction, such as the MiniRocket transform.}

   \item {We conduct a systematic review of existing CSI datasets, their characteristics, and experimental setups to enable reproducible research and standardized evaluation.}
   
    \item We present a qualitative and quantitative review of DL approaches, including both contrastive and non-contrastive SSL methods in Wi-Fi sensing, with experimental evaluations of selected approaches (SimCLR, VICReg, Barlow Twins, SimSiam) on complex scenarios including multi-user activity recognition and cross-domain adaptations. 

    Our experimental evaluation demonstrated that SSL methods achieved performance comparable to supervised learning in same-domain tasks, particularly with few-shot learning setups. However, task adaptation scenarios posed challenges due to limitations such as the quality of data, differences in environments, and task-specific variations, highlighting the need for more robust SSL approaches.
    
   \item We highlight current challenges and opportunities in the domain of Wi-Fi sensing, including the poor generalization of existing methods, the scarcity of accessible datasets and their limitations, the design of lightweight models, and the integration of multi-modal learning.
\end{itemize}

    
The structure of this paper is organized as follows: In Section~\ref{sec:preliminaries}, we present an overview of Wi-Fi standards, introduce the foundational concepts of Wi-Fi CSI, outline the methods used to measure CSI, and explore the impact of mobile objects on CSI in theory. In Section~\ref{sec:datasets}, we review and compare the available tools for CSI collection, provide a practical example of a CSI data measurement set-up for Wi-Fi sensing applications, and then present an overview of the available CSI datasets. Section~\ref{sec:preprocessing} describes CSI preprocessing techniques and algorithmic methods for enhancing CSI quality and obtaining more informative representations, while Section~\ref{sec:dl} explores DL approaches in Wi-Fi sensing, showcasing various architectures and strategies. Section~\ref{sec:SSL} focuses on SSL methods, detailing how different contrastive and non-contrastive techniques can be utilized for Wi-Fi sensing. Next, Section~\ref{sec:experiments} outlines the experimental setup used to evaluate both SSL algorithms and supervised baselines on three different datasets. Section~\ref{sec:challenges} discusses emerging challenges and opportunities for future research. Finally, we present our conclusions in Section~\ref{sec:conclusion}.

\section{Fundamentals of Wi-Fi Sensing}
\label{sec:preliminaries}
This section outlines the evolution of Wi-Fi standards, discusses preliminaries of CSI representation in stationary and mobile scenarios, and defines CSI measurement techniques. 

\subsection{Evolution of Wi-Fi Standards}
The journey of Wi-Fi began with the IEEE 802.11 standard in 1997, operating at 2.4 GHz with data rates up to 2~Mbps, laying the foundation for wireless communication. The subsequent introduction of IEEE 802.11b (Wi-Fi~1) and IEEE 802.11a (Wi-Fi~2) in 1999 brought enhanced speeds—up to 11 Mbps at 2.4~GHz with 802.11b and up to 54~Mbps at 5~GHz with 802.11a improving accessibility and reliability. By 2003, IEEE 802.11g (Wi-Fi~3) maintained the 54~Mbps data rate at 2.4~GHz, with improved range and compatibility.

\begin{table*}[h]
\scriptsize
\centering
\caption{Existing Survey Articles on Wi-Fi Sensing in the Last Five Years}
\begin{tabular}{M{0.8cm}M{1.3cm}M{3.5cm}M{3cm}M{4.5cm}M{3.8cm}}
\toprule
\textbf{Ref.} & \textbf{Focus} & \textbf{Signal Proc.\ Tech.} & \textbf{Models/DL Tech.} & \textbf{Key Contribution} & \textbf{Limitations} \\
\midrule
\cite{8863734} 2019 & ML & Feature extraction, denoising (smoothing filter, LP filter, PCA, linear interpolation, KF, WT) & DTW, SVM, iterative clustering, KNN, CNN & \begin{itemize}[leftmargin=*]\item IoT human sensing taxonomy: vital signs, gestures, activity recognition\item IoT sensing challenges analysis\end{itemize} & \begin{itemize}[leftmargin=*] \item No empirical validation \item Little on CSI preprocessing, ML models and DL architectures \item Dataset review is cursory \end{itemize} \\
\midrule

\cite{10.1145/3310194} 2019 & ML & Phase offset removal, denoising (Hampel, LOF), transforms (FFT, STFT, DWT), filtering \& thresholding & SVM, KNN, CNN, RNN, DTW & \begin{itemize}[leftmargin=*]\item First systematic Wi-Fi sensing categorization (detection, recognition, estimation)\item Signal processing framework and hybrid approaches analysis \end{itemize} & \begin{itemize}[leftmargin=*] \item No evaluation using state-of-the-art ML or DL models \item Lacks a comprehensive review of existing datasets \item Does not provide details on Wi-Fi CSI acquisition frameworks \end{itemize} \\
\midrule

\cite{9076313} 2020 & ML & Denoising (LP filter, BW filter), outlier removal, interpolation, T-F analysis, phase calibration & SVM, CNN, KNN, RF, LSTM & \begin{itemize}[leftmargin=*]\item Wi-Fi Vision applications review for smart environments\item Advanced human-centered sensing analysis \end{itemize} & \begin{itemize}[leftmargin=*] \item No coverage of publicly available Wi-Fi sensing datasets \item Does not examine the generalization capability of existing methods \end{itemize} \\
\midrule

\cite{9900419} 2022 & ML& Feature extraction (amplitude, phase, temporal difference, WT, PSD, statistics), denoising (window-statistics, S-G, Hampel, BW, DWT, FFT), dimensionality reduction (subset statistics, PCA, ICA) & KNN, DTW, CNN, SVM, DNN, GAN, LSTM, TPE & \begin{itemize}[leftmargin=*]\item In-depth exploration of preprocessing techniques and their impact on performance\item Edge-device Wi-Fi sensing with ESP32\item Resource-constrained validation \end{itemize} & \begin{itemize}[leftmargin=*]\item Limited examination of how the method generalizes across different conditions
\item Focused on ESP32; lacks detailed coverage of other CSI acquisition frameworks
\item Lacks a comprehensive review of existing datasets\end{itemize} \\
\midrule

\cite{yang2023sensefi} 2023 & DL & --- & MLP, CNN, RNN variants & \begin{itemize}[leftmargin=*]\item Transfer \& unsupervised learning \end{itemize} & \begin{itemize}[leftmargin=*] \item Limited coverage of CSI pre-processing techniques \item Review includes only a small set of datasets \item Lacks evaluation with state-of-the-art deep learning models \end{itemize} \\
\midrule

\cite{10.1145/3570325} 2023 & Few-shot learning & Domain-invariant feature extraction & Transfer learning, few-shot learning & \begin{itemize}[leftmargin=*]\item Cross-domain Wi-Fi sensing adaptability\item Algorithm limitations in domain shifts \end{itemize} & 
\begin{itemize}
    \item Some algorithms assume prior domain knowledge, such as user location or antenna placement, limiting general applicability.
    \item Lacks cross-dataset benchmarking; most evaluations are confined to within-dataset domain shifts.
\end{itemize}\\
\midrule

\cite{xu2023selfsupervisedlearningwificsibased} 2023 & SSL & Feature extraction (amplitude, phase), dimensionality reduction (mean-pooling) & SSL (SimCLR, MoCo, SwAV, Rel-Pos, MAE) & \begin{itemize}[leftmargin=*]\item First systematic evaluation of SSL algorithms on multiple CSI-HAR datasets\item Assessment of four SSL categories\item Development of performance metrics for real-world SSL applications \end{itemize} & \begin{itemize}[leftmargin=*] \item Only a limited number of datasets are examined \item Few details are provided on the CSI acquisition tools and Wi-Fi standards used \end{itemize} \\
\midrule

\cite{ahmadiftikhar2024wifi} 2024 & DL & CSI/RSSI, AoA, Doppler, FMCW & MLP, CNN, ResNet, RNNs, Transformers, AE, Hybrid & \begin{itemize}[leftmargin=*]\item First RIS integration for Wi-Fi sensing\item Multi-modal taxonomy for RF-VLC systems \end{itemize} & \begin{itemize}[leftmargin=*] \item No empirical validation \item Lack of practical guidelines for deploying solutions in real-world settings \item Limited examination of how methods generalize across different environments \end{itemize} \\
\midrule

\cite{xu2025evaluating} 2025 & SSL & Feature extractions (amplitude, phase), Doppler, ToF & SSL (SwAV, MoCo, MAE, Rel-Pos, SCN, SemiAMC) & \begin{itemize}[leftmargin=*] \item Comprehensive evaluation of SSL algorithms for CSI-based HAR \item Analysis across user, room, and receiver domain shifts (intra-dataset) \item Evaluation of SSL robustness under limited labeled data conditions \end{itemize} & \begin{itemize}[leftmargin=*] \item Focus only on SSL pre-training and fine-tuning; no investigation of downstream few-shot learning adaptation \item Cross-domain evaluation limited to user, room, and receiver shifts within the same dataset \item Lacks implementation guidelines for real deployment scenarios \end{itemize} \\
\midrule

\textbf{This Survey} & --- & \textbf{Denoising (DWT, FFT, Hampel), dimensionality reduction (PCA), CSI extraction} & \textbf{SSL (SimCLR, VICReg, Barlow Twins, SimSiam)} & \begin{itemize}[leftmargin=*]\item \textbf{Comprehensive analysis from data collection to CSI processing and DL techniques} \item \textbf{SSL performance evaluation on multi-user, multi-label tasks} \item \textbf{Cross-task domain-adaptation analysis} \item \textbf{Review of public datasets and evaluation standards} \end{itemize} & --- \\
\bottomrule
\end{tabular}
\label{rel_works1}
\footnotesize
\vspace{0.1cm}

\noindent \textit{Abbreviations:} AE: Autoencoder, ANN: Artificial Neural Network, AoA: Angle of Arrival, AoD: Angle of Departure, BW: Butterworth, CSI: Channel State Information, DWT: Discrete Wavelet Transform, FFT: Fast Fourier Transform, GRU: Gated Recurrent Unit, HAR: Human Activity Recognition, ICA: Independent Component Analysis, KF: Kalman Filter, LP: Low-Pass, MA: Moving Average, PCA: Principal Component Analysis, RF: Random Forest, RNN: Recurrent Neural Network, S-G: Savitzky–Golay, SSL: Self-Supervised Learning, TL: Transfer Learning, ToF: Time of Flight, ViT: Vision Transformer, WT: Wavelet Transform
\end{table*}

\twocolumn
\normalsize

A major leap occurred in 2009 with IEEE 802.11n (Wi-Fi 4), which supported both 2.4 GHz and 5 GHz frequencies and introduced MIMO technology, enabling speeds up to 600 Mbps. MIMO's ability to enhance capacity and reliability was crucial for the development of environmental mapping in Wi-Fi sensing. In 2013, IEEE 802.11ac (Wi-Fi 5) advanced Wi-Fi sensing by achieving data rates up to 3.5 Gbps at 5 GHz, enabling higher resolution for detailed motion and presence detection. The 2019 release of IEEE 802.11ax (Wi-Fi 6) further optimized performance in high-density environments, reaching speeds up to 9.6 Gbps across 2.4 GHz and 5 GHz bands. The extension Wi-Fi 6E, introduced in 2020, expanded these capabilities to the 6 GHz band, offering increased capacity and reduced congestion.

As Wi-Fi technology evolved, standards-based technologies emerged to bring an intelligent approach and interoperability across networks. {The IEEE 802.11k standard enables access points and clients to exchange information about the Wi-Fi environment, allowing devices to optimize their connections}. {IEEE 802.11v uses this network information to influence client roaming decisions, facilitating improved network management}, while {IEEE 802.11u enables devices to gather information from other networks before connecting}. {IEEE 802.11r further enhances network efficiency by enabling fast transitions between access points within a Wi-Fi network, supporting applications that require seamless mobility}. Together, these standards support Wi-Fi Agile Multiband, a technology that improves network management and interoperability across various vendors. Additionally, Wi-Fi Alliance-defined technologies supplement these standards by exchanging supplementary information, and identifying preferred channels, bands, or access points to further enhance intelligent Wi-Fi network management~\cite{banerji2013ieee}.
The  introduction of IEEE 802.11bf  in 2020 marked a significant milestone by supporting sensing across multiple frequency bands, including 2.4 GHz, 5 GHz, 6 GHz, and 60 GHz. This standard builds on the protocols of IEEE 802.11ad and IEEE 802.11ay for the 60 GHz band, minimizing communication overhead and making Wi-Fi sensing more accessible and efficient \cite{du2024overview}. 
Further advancements are expected with the forthcoming IEEE 802.11be (Wi-Fi 7), which aims to deliver data rates up to 30 Gbps across multiple frequency bands. With enhancements like multi-link operation and wider channel bandwidths. Wi-Fi 7 is poised to support even more accurate environmental data collection for advanced sensing applications.
Beyond Wi-Fi 7, the next major evolution—Wi-Fi 8, based on IEEE 802.11bn and anticipated by 2028—will operate across the 2.4, 5, and 6 GHz bands, introducing multiple access point coordination and transmission over millimetre wave (mmWave) frequencies. To reach speeds up to 100 Gbps, far surpassing copper Ethernet’s 40 Gbps limit, Wi-Fi 8 promises low-latency connectivity but may experience signal attenuation in rainy conditions, similar to satellite communications. Deploying Wi-Fi 8 will require retrofitting ceiling-mounted access points to support these advancements \cite{ieee_evolution_WiFi,galati2024will}.

{The IEEE 802.11bf standard represents a pivotal advancement, formally integrating Wi-Fi sensing capabilities such as gesture recognition, presence detection, and environmental monitoring into mainstream wireless protocols, while ensuring coexistence with legacy Wi-Fi deployments~\cite{ropitault2024ieee, keshtiarast2025next}. This amendment bridges the gap between Wi-Fi sensing research and practical deployment, standardizing features for bistatic and multistatic sensing across the 2.4 GHz, 5 GHz, 6 GHz, and 60 GHz bands. Building upon this foundation, Wi-Fi 7 (802.11be) introduces wider bandwidths, multi-link operation (MLO), and lower latencies, enabling more accurate and real-time environmental sensing~\cite{liu2024wi}. Looking ahead, Wi-Fi 8 (802.11bn) is poised to revolutionize high-demand sensing applications by prioritizing ultra-high reliability (UHR) and integrating millimeter-wave (mmWave) communications~\cite{galati2024will, liu2024wi}. These enhancements will enable future sensing use cases such as augmented reality (AR) and holographic communications, where immediate feedback, extremely high throughput, and minimal latency are critical~\cite{hatami2024survey}. Wi-Fi 8’s capabilities will not only enhance conventional sensing but also open new avenues for immersive applications in the metaverse and real-time digital twin systems, overcoming many hardware limitations that constrained Wi-Fi 7.}

\subsection{Preliminaries of Channel State Information (CSI)}
In a MIMO-OFDM\footnote{MIMO  employs multiple antennas at both the transmitter and receiver ends, creating a matrix of connections between every transmitter and receiver. Orthogonal Frequency Division Multiplexing (OFDM) divides the channel bandwidth into multiple smaller, non-overlapping frequency bands, called subcarriers, which transmit bit streams in parallel. } system, the received signal \( \mathbf{y} \) can be modeled as follows:
\begin{equation}
\mathbf{y} = \mathbf{H} \mathbf{x} + \boldsymbol{\eta},
\end{equation}
where \( \mathbf{x} \) is the transmitted signal vector, \( \boldsymbol{\eta} \) is the additive noise vector, and \( \mathbf{H} \) is the CSI matrix, capturing the properties of the wireless  channels between the transmitter and receiver.
Each element \(H_{m,n}(f_c; t) \) in \( \mathbf{H} \) represents the channel impulse response (CIR) between the \( n \)-th transmitter antenna and the \( m \)-th receiver antenna at a given frequency \( f_c \) and time \( t \), encapsulating environmental effects such as path loss, obstacles, and interference. The CSI matrix element \( H_{m,n}(f_c; t) \) can be expressed in terms of the CIR as follows:
\begin{equation}
H_{m,n}(f_c; t) = \sum_{l=1}^{L} a_{l}(t) e^{-j2\pi f_c \tau_{l}(t)},
\end{equation}
where \( L \) is the number of multipath components, \( a_{l}(t) \) represents the amplitude attenuation, and \( \tau_{l}(t) \) is the propagation delay for each path.
The amplitude \( \lvert H_{m,n}(f_c; t) \rvert \) and phase \( \angle H_{m,n}(f_c; t) \) of CSI provide detailed information about multipath propagation. The structure of the CSI matrix at a given receiver depends on the number of receiving and transmitting antennas, subcarriers, and timestamps in the wireless communication system. It can be visualized as a multidimensional array, where each element provides detailed channel information for a specific combination of time, subcarrier, transmit antennas, and receive antennas. Wi-Fi devices obtain accurate CSI measurements by utilizing \textit{Long Training Fields} (LTFs). By comparing the received LTFs with the known transmitted patterns, the receiver can estimate the CSI, effectively producing a ``Wi-Fi image'' of the surrounding environment \cite{Yang_Chen_Wang_Zou_Lu_Sun_Xie_2022}. 

\begin{table*}[h!]
\centering
\small
\caption{Evolution of Wi-Fi Standards and Their Specifications}
\label{comparison_table}
\resizebox{\textwidth}{!}{%
\begin{tabularx}{\textwidth}{M{1.8cm} M{0.8cm} M{2.0cm} M{2.0cm} M{1.5cm} M{7.4cm}}
\toprule
\textbf{IEEE Standard} & \textbf{Year} & \textbf{Frequency Bands} & \textbf{Channel Width} & \textbf{Data Rate} & \textbf{Notes} \\ 
\midrule
802.11 (Wi-Fi 0) & 1997 & 2.4 GHz & 20 MHz & Up to 2 Mbps & 
\begin{itemize}[nosep,leftmargin=*]
    \item Laid the foundation for wireless communication.
\end{itemize} \\ 
\midrule
802.11b (Wi-Fi 1) & 1999 & 2.4 GHz & 20 MHz & Up to 11 Mbps & 
\begin{itemize}[nosep,leftmargin=*]
    \item Increased data rates, making Wi-Fi more widely adopted.
\end{itemize} \\ 
\midrule
802.11a (Wi-Fi 2) & 1999 & 5 GHz & 20 MHz & Up to 54 Mbps & 
\begin{itemize}[nosep,leftmargin=*]
    \item Operated in 5 GHz, providing higher data rates and less interference.
\end{itemize} \\ 
\midrule
802.11g (Wi-Fi 3) & 2003 & 2.4 GHz & 20 MHz & Up to 54 Mbps & 
\begin{itemize}[nosep,leftmargin=*]
    \item Combined speed of 802.11a with 802.11b's range and compatibility.
\end{itemize} \\ 
\midrule
802.11n (Wi-Fi 4) & 2009 & 2.4 GHz and 5 GHz & 40 MHz & Up to 600 Mbps & 
\begin{itemize}[nosep,leftmargin=*]
    \item Introduced MIMO for improved speed and reliability.
    \item Enabled better performance on both 2.4 GHz and 5 GHz bands.
\end{itemize} \\ 
\midrule
802.11ad (WiGig) & 2012 & 60 GHz & 2.16 GHz & Up to 8 Gbps & 
\begin{itemize}[nosep,leftmargin=*]
    \item Utilized 60 GHz for short-range, high-speed, low-latency applications.
\end{itemize} \\ 
\midrule
802.11ac (Wi-Fi 5) & 2013 & 5 GHz & 80/160 MHz & Up to 3.5 Gbps & 
\begin{itemize}[nosep,leftmargin=*]
    \item Enhanced with wider channels, more channels, and higher modulation density.
\end{itemize} \\  
\midrule
802.11ax (Wi-Fi 6) & 2019 & 2.4 GHz and 5 GHz & Up to 160 MHz & Up to 9.6 Gbps & 
\begin{itemize}[nosep,leftmargin=*]
    \item Optimized for dense environments with OFDMA and better spectral efficiency.
\end{itemize} \\ 
\midrule
802.11ax (Wi-Fi 6E) & 2020 & 6 GHz & Up to 160 MHz & Up to 9.6 Gbps & 
\begin{itemize}[nosep,leftmargin=*]
    \item Extended Wi-Fi 6 into the 6 GHz band for increased capacity and reduced congestion.
\end{itemize} \\ 
\midrule
802.11bf & 2020 & 2.4 GHz, 5 GHz, 6 GHz, and 60 GHz & N/A & N/A & 
\begin{itemize}[nosep,leftmargin=*]
    \item Focused on Wi-Fi sensing, enabling applications in motion and presence detection.
\end{itemize} \\ 
\midrule
802.11ay & 2021 & 60 GHz & 2.16/4.32/6.48/\hspace{0.01cm}64 GHz & Up to 176 Gbps & 
\begin{itemize}[nosep,leftmargin=*]
    \item Enhanced 802.11ad with higher data rates and channel bonding.
\end{itemize} \\ 
\midrule
802.11be (Wi-Fi 7) & 2024 & 2.4 GHz, 5 GHz, and 6 GHz & Up to 320 MHz & Up to 30 Gbps & 
\begin{itemize}[nosep,leftmargin=*]
    \item Aims for higher data rates with multi-link operation and wider channels.
\end{itemize} \\ 
\midrule
{802.11bn (Wi-Fi 8)} & {2028} & {2.4 GHz, 5 GHz, and 6 GHz} & {TBD} & {Up to 100 Gbps} & 
\begin{itemize}[nosep,leftmargin=*]
    \item Utilizes multiple access point coordination and mmWave frequencies.
    \item Optimized for low-latency applications and very high data rates.
    \item Rain attenuation similar to satellite communications may impact performance.
    \item Retrofit of ceiling-mounted access points required for deployment.
\end{itemize} \\ 
\bottomrule
\end{tabularx}
}

\small
\vspace{0.1cm}
\noindent \textit{Abbreviations:} MIMO: Multiple-Input Multiple-Output, OFDMA: Orthogonal Frequency-Division Multiple Access, NDPs: Null Data Packets
\end{table*}

\begin{figure}[h]
\centering
\includegraphics[width=0.5\textwidth]{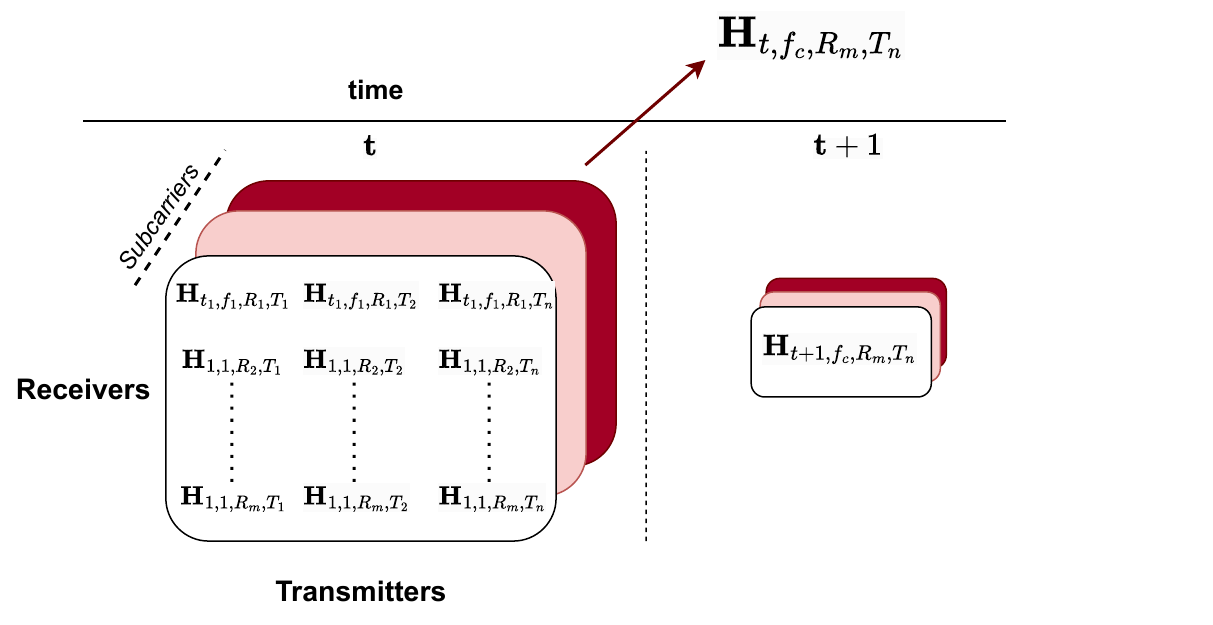}
\caption{Structure of the CSI matrix, illustrating dimensions across receivers, transmitters, subcarriers, and time.}
\label{fig:CSI_Matrix_Structure}
\end{figure}

As shown in \textbf{Fig.~\ref{fig:CSI_Matrix_Structure}}, the hierarchical structure of the CSI matrix is depicted, illustrating how it evolves and frequency. At each time step \( t \), the matrix \( H_{t,f_c,R_m,T_n} \) captures wireless channel measurements between multiple transmit antennas \( T_n \) and receive antennas \( R_m \) across different subcarrier frequencies \( f_c \). Each element of the matrix represents a complex channel coefficient containing both amplitude and phase information, characterizing the wireless channel between a specific transmitter-receiver pair at a specific frequency and time.

\textbf{Fig. \ref{fig:CSI_Matrix_Structure}} demonstrates that the CSI matrix can be thought of as layers stacked over time and frequency. Each layer corresponds to a different subcarrier frequency, and within each layer, the grid formed by the receivers and transmitters provides a detailed snapshot of the channel characteristics at that specific time and frequency. Although the figure presents an example with two time instances, this structure can extend across any number of timestamps and subcarriers.

\subsection{Active and Passive CSI Measurements}

\begin{figure}
\centering
\begin{subfigure}{1.0\linewidth}
    \centering
    \includegraphics[width=0.58\linewidth]{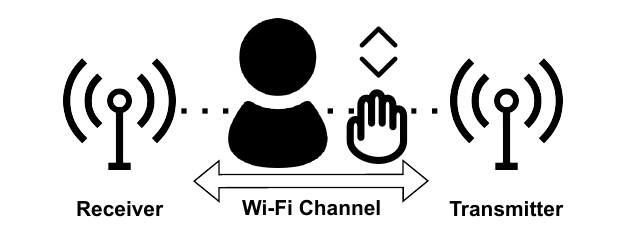}
    \caption{Active Wi-Fi Sensing}\label{fig:image1}
\end{subfigure}

\bigskip
\begin{subfigure}{\linewidth}
  \centering
    \includegraphics[width=0.80\linewidth]{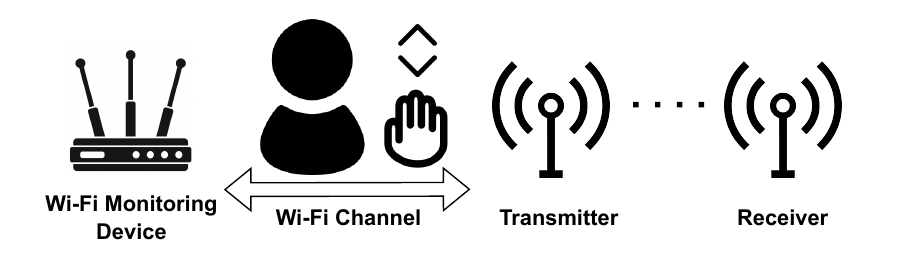}
  \caption{Passive Wi-Fi Sensing}\label{fig:image3}
\end{subfigure} 
{\caption{Active sensing and passive sensing setups for CSI data collection.}
\label{fig:activepassive}}
\end{figure}

Collecting CSI data involves configuring transmitters and receivers, with at least one Wi-Fi network interface card (NIC) running modified firmware to support CSI extraction. This can be done using either active or passive sensing methods, depending on the application and setup.
In active sensing, a transmitter sends predefined packets to a receiver, which extracts and stores CSI data. This approach requires only two devices and offers precise control over the signal, as the devices’ positions are known and packet rates can be adjusted for greater accuracy. Active sensing is ideal for scenarios needing consistent and predictable signal patterns. However, it consumes bandwidth, rendering the transmitter unavailable for other tasks. \textbf{Fig.~\ref{fig:activepassive}(a)} illustrates an active Wi-Fi sensing setup, where CSI captures the channel between the transmitter and receiver.

{Passive sensing, on the other hand, leverages existing Wi-Fi traffic without introducing additional transmissions. A monitoring device collects CSI from ongoing communications and estimates the channels to transmitters with known MAC addresses. This setup typically involves at least three devices. Passive sensing minimizes interference and is suitable for scenarios where active transmission is impractical or disruptive.} {Passive sensing is advantageous due to its lower bandwidth usage and reduced power consumption, making it ideal for applications like HAR and smart environments where non-intrusiveness is critical. However, it poses challenges, such as the need for precise device synchronization, lower packet rates, and the possibility of missed packets due to dependence on existing traffic. \textbf{Fig.~\ref{fig:activepassive}(b)} shows a passive Wi-Fi sensing configuration, where CSI captures the channel between a monitoring device and a transmitter.}

\subsection{Wi-Fi CSI for Moving Users}
\begin{figure}[b]
        \centering		\includegraphics[width=0.80\linewidth]{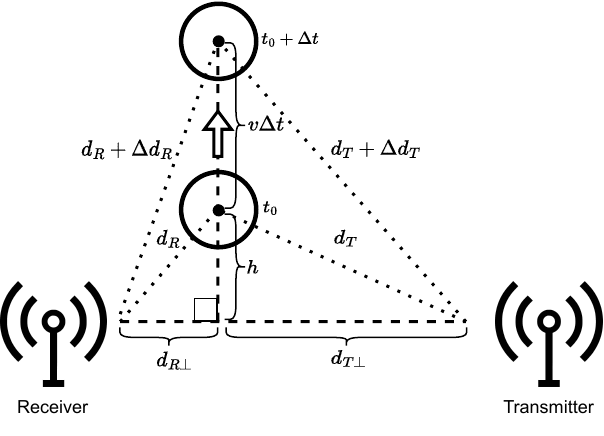}	
	\caption{The spatial arrangement of a Wi-Fi transmitter and receiver  where an object moves upward at a velocity of $v$ \cite{10403438}.}
	\vspace{-4mm}
\label{figure:CSI_moving}
\end{figure}

To understand the effect of a moving point, such as a human hand, on the Wi-Fi CSI phase, we can consider a scenario in which an object is positioned between a Wi-Fi transmitter and receiver, see \textbf{Fig. \ref{figure:CSI_moving}}. As the object moves upward at a constant velocity \(v\) over a duration of \(\Delta t\) seconds, the altered signal path introduces a time delay \(\Delta \tau\), which in turn causes a phase shift. Accounting for this delay, the CSI \(h(f;t)\) for a subcarrier with carrier frequency \(f_c\) at time \(t + \Delta t\) can be expressed as follows:
\begin{equation}
\begin{aligned}
h\left(f_c; t + \Delta t\right) & =h\left(f_c; t\right) \exp \left(-j 2 \pi f_c\Delta \tau\right),
\end{aligned}
\label{equation:6}
\end{equation}
\noindent
in which $\Delta \tau$ is
\begin{equation}
\begin{aligned}
\Delta \tau & =\frac{\Delta d(v)}{c},
\end{aligned}
\label{equation:7}
\end{equation}
where \(c\) represents the propagation speed of Wi-Fi signals through the air and \(\Delta d(v)\) corresponds to the displacement resulting from the object's motion \cite{10403438}.

If there are $L$ moving points in the space that reflect the transmitted signal to the receiver, the resulting CSI signal can be expressed as follows
\begin{equation}
\begin{aligned}
\resizebox{0.8\hsize}{!}{$
h\left(f_c; t_0\! +\! \Delta t\right)\! =\!\sum_{i=1}^L h_i\left(f_c; t_0\right) \exp\! \left(\!-j 4 \pi f_c(\frac{v_i\Delta t}{c})\right),$}
\end{aligned}
\label{equation:9}
\end{equation}
\noindent
where $h_i(f_c; t)$ represents the CSI from the $i^{\mathrm{th}}$ path at time $t$ and carrier frequency $f_c$, while $v_i$ denotes the velocity of the $i^{\mathrm{th}}$ object. Suppose that $M$ samples of CSI are obtained with a sampling interval of $\Delta t$, and the instantaneous velocities of the moving points are constant. Under these conditions, \textbf{Eq.~\ref{equation:9}} for the CSI vector $\mathbf{h}_{f_c}$, consisting of $M$ elements, can be represented in matrix form as follows:
\begin{equation}
\begin{aligned}
\mathbf{h}_{f_c}\!=\!
\underbrace{\left[\mathbf{d}\!\left(v_1\right)\!, \!\ldots\!, \mathbf{d}\left(v_L\right)\right]}_{\triangleq \mathbf{D}_{M \times L}}\!
\underbrace{\left[h_1\!\left(t_0, \!f_c\right)\!, \!\ldots\!, h_L\!\left(t_0,\! f_c\right)\right]^{\top}}_{\triangleq \mathbf{a}_{L \times 1}},
\end{aligned}
\label{equation:10}
\end{equation}
\noindent
where, for each moving point indexed by \(i\), the velocity vector \(\mathbf{d}(v_i)\), composed of \(M\) elements, is given by
\begin{equation}
\begin{aligned}
\mathbf{d}(v_i)\!=\!\left[1, e^{-j 4 \pi f_c(\frac{v_i\Delta t}{c})}\!,\!\dots, e^{-j 4 \pi (M-1) f_c(\frac{v_i\Delta t}{c})}\right],
\end{aligned}
\label{equation:11}
\end{equation}
\noindent
Thus, $\mathbf{D}$ represents the $M\times L$ matrix of velocity vectors. Thus, considering a random noise vector $\mathbf{n}$ with $M$ elements, \textbf{Eq.~\ref{equation:10}} can be reformulated as follows:
\begin{equation}
\begin{aligned}
\mathbf{h}_{f_c} = \mathbf{D}\mathbf{a} + \mathbf{n}.
\end{aligned}
\label{equation:12}
\end{equation}
where $\mathbf{a}$ represents the vector of initial CSI at the reference time $t_0$. The equation above establishes a relationship between the collected CSI samples and the velocity of objects moving in the environment. This suggests that velocities, referred to as Doppler velocity, can be estimated from CSI samples \(\mathbf{h}_{f_c}\), thus relating CSI values to the human activities performed within the environment.

\section{CSI Extraction Tools and Existing Datasets}
\label{sec:datasets}

In this section, we present a review of the standard CSI extraction tools, pin-point existing CSI datasets and their specifications, and discuss a simplified testbed for CSI extraction. 


\begin{table*}[h]
\centering
\caption{Comparison of CSI Tools}
\label{comparison_table}
\resizebox{\textwidth}{!}{
\begin{tabularx}{\textwidth}{M{2.5cm} M{3.0cm} M{1.5cm} M{3.5cm} M{3.0cm} M{1.5cm}}
\toprule
\textbf{Tool Name}  & \textbf{Hardware} & \textbf{Antennas} & \textbf{Subcarriers (Channel Bandwidth)} & \textbf{Independent Functionality} & \textbf{Cost (CAD)} \\ 
\midrule

802.11n CSI Tool & Intel 5300 NIC & 3 & 56 (20 MHz) / 114 (40 MHz) & No & \$20 + PC \\ 

Atheros CSI Tool & Atheros NIC & 3 & 56 (20 MHz) / 114 (40 MHz) & No & \$20 + PC \\ 

Nexmon CSI Tool & Broadcom Wi-Fi Chips & 1-4 & 64 (20 MHz) / 128 (40 MHz) / 256 (80 MHz) & Yes (firmware required) & \$130 \\ 

ESP32 CSI Tool & ESP32 Microcontrollers & 1 & 64 (20 MHz) & Yes & \$14 \\ 

WARP Tool & Custom FPGA Platform & Configurable & Configurable across multiple bands & Yes & \$500+ \\ 
\bottomrule

\end{tabularx}
}
\par
\vspace{0.1 cm}
\small \textit{Notes: PC = Personal Computer, CSI = Channel State Information, NIC = Network Interface Card, FPGA = Field Programmable Gate Array.}
\end{table*}

\subsection{CSI Extraction Tools}

CSI extraction is critical since the IEEE 802.11n standard requires specialized tools, as not all Wi-Fi cards are equipped to report CSI. This section reviews prominent tools used for CSI extraction, thus emphasizing their technical capabilities and suitability for various Wi-Fi sensing applications.

\subsubsection{802.11n CSI Tool}
The 802.11n CSI tool \cite{halperin2011tool} is widely recognized for using Intel 5300 Wi-Fi cards, capturing CSI data across 30 groups of subcarriers for each antenna pair. In a 20 MHz channel, it measures 56 total subcarriers, while in a 40 MHz channel, it captures 114 subcarriers. This tool is popular for its reliability in obtaining detailed and compressed CSI, making it suitable for numerous research applications in Wi-Fi-based sensing.

\subsubsection{Atheros CSI Tool}
The Atheros CSI Tool \cite{10.1145/2789168.2790124}, designed for use with Qualcomm Atheros cards, records uncompressed CSI data. This approach improves resolution by capturing individual CSI values for all 56 subcarriers within a 20 MHz channel and 114 subcarriers in a 40 MHz channel \cite{csiaethtool}. Unlike tools that compress CSI, the Atheros CSI Tool retains the full precision of amplitude and phase information, providing finer granularity and enabling high-resolution analysis of the wireless channel. This level of detail is particularly advantageous for advanced applications in Wi-Fi sensing and channel modeling.

\subsubsection{Wireless Open Access Research Platform (WARP)}
The Wireless Open Access Research Platform (WARP) \cite{10.1145/3310194} extends CSI measurement capabilities across multiple frequency bands. WARP is designed for versatility and allows for detailed analysis across a broader spectrum, meeting advanced requirements for multi-band Wi-Fi sensing.

\subsubsection{Nexmon CSI Tool}
The Nexmon CSI Tool \cite{schafer2021human} enables CSI extraction on devices such as the Raspberry Pi and selected router models, capturing data on all 256 subcarriers in 80 MHz channels. For 40 MHz channels, it captures 128 subcarriers, including 108 data subcarriers, 6 pilot subcarriers, and 14 null subcarriers. While this data can be noisier than other tools, Nexmon’s compatibility with accessible and affordable hardware makes it a practical solution for Wi-Fi-based monitoring in cost-sensitive applications~\cite{sharma2021passive}.

\subsubsection{ESP32 CSI Tool}
The ESP32 CSI Tool \cite{Hern2006:Lightweight} provides an efficient solution for CSI data collection on ESP32 microcontrollers. Operating on a 20 MHz channel, it captures CSI data for 64 subcarriers, offering sufficient granularity for Wi-Fi sensing applications. Supporting stand-alone operation, this tool is ideal for large-scale deployments of Wi-Fi sensing, making it a suitable choice for Internet of Things (IoT) and resource-constrained environments.

Key factors in comparing CSI extraction tools include cost, complexity, compatibility, and independent functionality—whether the tool can operate autonomously or requires additional hardware or software support. As summarized in \textbf{Table~\ref{comparison_table}}, the 802.11n CSI Tool (Intel 5300) and the Atheros CSI Tool (Qualcomm Atheros) remain popular choices for CSI extraction, but their reliance on specific Wi-Fi cards and the need for connection to external devices such as laptops or desktops increase their operational complexity and cost.
In contrast, the Nexmon CSI Tool (Broadcom) demonstrates greater flexibility with its broader compatibility and the ability to operate independently through firmware modifications, particularly on devices like the Raspberry Pi. This makes Nexmon a scalable and practical option for cost-sensitive or portable applications.
Finally, the ESP32 CSI Tool emerges as the most portable and cost-effective solution. Its ability to operate independently on a lightweight ESP32 microcontroller and to store CSI data directly on an onboard micro SD card eliminates the need for external devices, enabling seamless deployment in diverse scenarios. However, the lower quantization level of the ESP32 CSI tool compared to other tools may limit its sensitivity for certain high-precision sensing applications, making it more suitable for use cases where affordability and portability take precedence over extreme precision.

\subsection{A Typical Cost-effective Passive CSI Measurement Set-up}
Wi-Fi CSI measurement requires carefully designed experiments to collect high-quality datasets. Key factors include the placement of Wi-Fi devices, environmental conditions, and participant behaviour to ensure the data is both representative and informative. 
In previous studies, a common passive CSI measurement setup involves using an ASUS RT-AC86U router with multiple external antennas as the access point and a Raspberry Pi 3B or 4B as the transmitter. A Wi-Fi modem ensures a stable connection between the Raspberry Pi and the network, enabling uninterrupted packet transmission. The ASUS router is typically chosen for its compatibility with Nexmon firmware modifications for CSI extraction, and its multiple antennas enhance measurement diversity.

CSI collection begins by selecting an unrestricted wireless channel (2.4 GHz or 5 GHz), as current frameworks can only collect packets from one channel at a time. A higher packet rate improves the sampling rate, which helps in CSI de-noising; a rate of at least 50 or 100 Hz is recommended for HAR applications \cite{8067693}. Packets can be sent to any arbitrary address from the transmitter. The ASUS router needs to be initialized with the Nexmon framework by specifying the wireless channel and the target transmitter's MAC address before starting monitoring. Afterwards, the CSI collection process can be initiated using the \texttt{tcpdump} command, storing packets in a \textit{.pcap} file. The embedded CSI values can later be extracted using tools like the CSIKit\footnote{\href{https://github.com/Gi-z/CSIKit}{https://github.com/Gi-z/CSIKit}} library for Python.

To capture a comprehensive understanding of user activity, multiple monitoring devices like additional ASUS routers can be employed. Synchronization between different CSI measurements is crucial in multi-device setups and can be achieved through post-processing algorithms. Additional sensors such as accelerometers, gyroscopes, or video cameras can assist in synchronization and CSI signal segmentation to ensure each data segment corresponds to a single user activity.

\subsection{Existing CSI Datasets}
Publicly available datasets play a vital role in advancing CSI-based Wi-Fi sensing research, offering diverse environments, hardware setups, and applications. These datasets generally fall into three categories: \textit{ HAR}, which targets broad, complex actions involving the whole body; \textit{Gesture Recognition and Sign Language}, focused on discrete, repetitive movements of specific body parts; and \textit{Health Monitoring and Specialized Datasets}, designed for applications like health tracking or unique environments. Most datasets consist of time-series data with labeled entries corresponding to activities or gestures. Typically, this data includes CSI  values, capturing signal amplitude and phase across multiple antennas and subcarriers, often supplemented by RSSI values and timestamps. In the following, we list the most popular datasets and their specifications. \textbf{Table~\ref{datasets_table}} provides a comparative summary of these datasets in terms of the hardware, extraction tool, and specifications.

\subsubsection{HAR Datasets}
In what follows, we organize datasets mainly designed for HAR applications.
\begin{itemize}
    \item \textbf{{UT-HAR Dataset:}}  
The UT-HAR dataset \cite{8067693} was collected in an indoor office environment using an Intel 5300 NIC with the Linux 802.11n CSI tool. It includes six activities (e.g., walking, sitting) performed by six individuals under LOS conditions. Data columns contain timestamps, 90 amplitude values across 30 subcarriers and 3 $R_m$ antennas operating at 5 GHz frequency, followed by 90 corresponding phase values, resulting in a CSI format of 1 x 250 x 90 with an approximate 4 GB size. Following the IEEE 802.11n/ac standard, this dataset, despite challenges with noise and variability, is valuable for developing activity recognition models and is available on \href{https://github.com/ermongroup/Wi-Fi_Activity_Recognition}{GitHub}.

\item \textbf{WiAR Dataset:}
The WiAR dataset \cite{8866726} comprises 16 activities performed by 10 volunteers across three indoor environments, collected with an Intel 5300 NIC operating on IEEE 802.11n standard at 5 GHz with 20 MHz bandwidth. It includes raw RSSI and CSI data across 30 subcarriers, 1 $T_n$ and 3 $R_m$ antennas with an approximate 5 GB size. Each row represents time-specific data consisting of two periods, active (human activity is being performed) and empty (background data with no activity). This structure helps in smart homes and healthcare environments by enabling activity recognition in a dynamic environment.

\item \textbf{{WiCount Dataset:}}
The WiCount dataset \cite{8367378} captures CSI data with detailed amplitude and phase information across 30 subcarriers, 2 $T_n$ and 3 $R_m$ antennas, collected using an Intel NUC laptop and mini R1C router operating on IEEE 802.11n standard at 5 GHz with 20 MHz bandwidth. This dataset allows volunteers free movement, enhancing its applicability in scenarios where precise movement tracking is essential. It is available on \href{https://github.com/wicount/WiCount/tree/master}{GitHub}.

\item \textbf{WiMANS Dataset:}
The WiMANS dataset \cite{huang2024wimans} is designed for multi-user activity sensing in smart homes and security contexts, capturing interactions of up to five users in dynamic environments. Using an Intel 5300 NIC operating on IEEE 802.11n standard with 3 $T_n$ transmitting and 3 $R_m$ receiving antennas, it collects CSI data across 30 subcarriers at both 2.4 GHz and 5 GHz frequencies. The dataset provides over 9.4 hours of CSI data synchronized with video recordings. The dataset, 1.44 GB in size, is accessible on \href{https://github.com/huangshk/WiMANS}{GitHub}.

\item \textbf{FallDeFi Dataset:}
FallDeFi \cite{10.1145/3161183} supports activity and fall detection applications. It includes CSI traces for activities such as falling, walking, and sitting, collected with Intel 5300 NICs operating on IEEE 802.11n standard at 5 GHz with 20 MHz bandwidth, using 2 $T_n$ transmitting and 2 $R_m$ receiving antennas across 30 subcarriers in typical indoor spaces (bedrooms, corridors). This time-sequence data is valuable for fall detection research and is available on \href{https://github.com/dmsp123/FallDeFi}{GitHub}.


\end{itemize}

\subsubsection{Gesture Recognition and Sign Language Datasets}
In what follows, we organize datasets mainly designed for gesture recognition and sign language applications.
\begin{itemize}
    \item {\textbf{Widar3.0 Dataset:}}
Introduced in 2019, the Widar3.0 dataset \cite{9516988} focuses on gesture recognition with CSI data collected from 16 users across various indoor environments using Intel 5300 NIC utilizing 1 $T_n$, and 18 $R_m$ across 30 subcarriers at 5.8 GHz. Its Body-coordinate Velocity Profile (BVP) feature enhances recognition accuracy in diverse settings, making it suitable for applications needing robust performance. The dataset is 3.4 GB and available online\footnote{\url{http://tns.thss.tsinghua.edu.cn/widar3.0/index.html}}.

\item {\textbf{SignFi Dataset:}}
The SignFi dataset \cite{10.1145/3191755} is the first to enable sign language recognition using  CSI. It includes 276 gestures from 5 users, captured in lab and home environments with a 5 ms sampling rate and noise preprocessing to ensure accuracy. Collected using Intel 5300 NIC following IEEE 802.11n at 5 GHz and 20 MHz frequency, utilizing 3 $T_n$ and 1 $R_m$ antennas. The dataset is 1.44 GB and accessible online\footnote{\url{https://yongsen.github.io/SignFi/}}.

\item \textbf{Wi-SL Dataset:}
The Wi-SL dataset \cite{s20144025} enables fine-grained gesture recognition through Wi-Fi CSI, without requiring wearable devices. Collected using TP-Link WDR4310 routers operating on IEEE 802.11n standard at 5 GHz with 40 MHz bandwidth, utilizing 2 $T_n$ transmitting and 3 $R_m$ receiving antennas across 114 subcarriers, with CSI extracted using the Atheros tool. It provides a non-invasive, cost-effective option suitable for smart home and assistive applications. Preprocessing may be necessary to mitigate challenges such as noise and multipath effects.

\item \textbf{CrossSense Dataset:}
The CrossSense Dataset \cite{10.1145/3241539.3241570} comprises 1,200,000 activity samples from 100 users performing 40 gestures, captured using both Intel 5300 NIC and TP-Link WDR7500 devices operating on IEEE 802.11n at 5 GHz. The system utilizes 3 $T_n$ transmitting and 3 $R_m$ receiving antennas across 30 subcarriers. Focused on gesture recognition and human-computer interaction, its robust design enables reliable performance across diverse scenarios and large-scale implementations. The dataset is available on \href{https://github.com/nwuzj/CrossSense}{GitHub}.

\item \textbf{DeepSeg Dataset:}
DeepSeg \cite{9235578} supports both fine-grained and coarse-grained HAR. Collected in a meeting room setting using a Linux-based Intel 5300 NIC operating on IEEE 802.11n standard with 1 $T_n$ transmitting and 3 $R_m$ receiving antennas across 30 subcarriers, it includes data for 10 activities performed by 5 volunteers, making it versatile for model evaluation. The dataset is available on \href{https://github.com/ChunjingXiao/DeepSeg}{GitHub}.
\end{itemize}

\subsubsection{Health Monitoring and Specialized Datasets}
In what follows, we organize datasets mainly designed for health monitoring applications.
\begin{itemize}

\item \textbf{eHealth CSI Dataset:}
The eHealth CSI dataset \cite{10177905} provides high-resolution CSI data combined with phenotype and heartbeat information for health monitoring applications. Using a Raspberry Pi 4B operating on IEEE 802.11n standard at 5 GHz with 80 MHz bandwidth, the system captures data using 1 $T_n$ transmitting and 1 $R_m$ receiving antenna across 234 subcarriers. Collected in a controlled setting, this dataset supports applications in physical analysis and activity recognition. Access requires a request form.

\item \textbf{IEEE 802.11ac 80 MHz CSI Dataset:}
This dataset \cite{10144501} includes channel measurements over an 80 MHz and 5 GHz bandwidth, suited for HAR and person identification. Using 3 $T_n$ and 3 $Rm$ antennas across 242 subcarriers have been collected across seven diverse environments, it offers time diversity and spans 23.6 GB. Available for download via \href{https://ieee-dataport.org/documents/csi-dataset-wireless-human-sensing-80-mhz-wi-fi-channels}{GitHub}.

\item \textbf{Wi-Fi CSI-based HAR Dataset:}
Supporting smart home automation and security, this dataset \cite{ZHURAVCHAK202259} captures activities in three rooms, with each activity labeled with an image and bounding box. Data collection utilized 2 TP-Link TL-WDR4300 routers with 3 $T_n$ transmitting and 3 $R_m$ receiving antennas, operating on dual bands: 2.4 GHz with 20 MHz bandwidth across 56 subcarriers, and 5 GHz with 40 MHz bandwidth across 114 subcarriers. The dataset is available on \href{https://github.com/Retsediv/WIFI_CSI_based_HAR}{GitHub} in two versions: 1.2 GB without images or 9.1 GB with images.

\item \textbf{CSI Human Activity Dataset:}
Collected in rooms of approximately 4 m × 4.5 m, this dataset \cite{app11198860} supports healthcare and smart living applications. Data collection employed Raspberry Pi 3B+ and 4B with ASUS RT-AC86U routers, utilizing 2 $T_n$ transmitting and 2 $R_m$ receiving antennas, with CSI extracted using the Nexmon tool. The system operates in two configurations: 128 subcarriers at 40 MHz bandwidth and 256 subcarriers at 80 MHz bandwidth. Although inherently noisy, this 23.6 GB dataset provides comprehensive data for activity recognition tasks and is available on \href{https://ieee-dataport.org/open-access/csi-human-activity}{GitHub}.

\item \textbf{WiADG Dataset:}
The WiADG dataset \cite{8487345} focuses on gesture recognition using device-free Wi-Fi, featuring two TP-Link TL-WDR4300 routers on IEEE 802.11n at 5 GHz with 40 MHz bandwidth. The setup incorporates 3 $T_n$ transmitting and 3 $R_m$ receiving antennas, spanning 114 subcarriers, with CSI extracted using the Atheros tool. Collected in two indoor environments, it is optimized for gesture recognition via adversarial domain adaptation and is accessible on \href{https://github.com/NTU-AIoT-Lab/WiADG}{GitHub}.
\end{itemize}

\subsubsection{Other Datasets}
\begin{itemize}
    \item \textbf{ZigFi Dataset:} ZigFi \cite{8970452} was collected using Intel 5300 NICs and TelosB motes in environments with varying noise, including offices during day and night. It is valuable for studying environmental impacts on CSI. The data is available on \href{http://tns.thss.tsinghua.edu.cn/sun/static/data/zigfi.zip}{GitHub}.

    \item \textbf{EyeFi Dataset:}
The EyeFi dataset \cite{9183685} combines vision data from a Bosch Flexidome IP Panoramic Camera with Wi-Fi data from an Intel 5300 NIC. Collected in lab and kitchen settings, it contains over 1.2 million Wi-Fi packets and is valuable for multi-modal research. Available on \href{https://zenodo.org/record/3882104}{Zenodo}.
\end{itemize}

\begin{table*}[]
\centering
\caption{Publicly Available Datasets}
\label{datasets_table}
\resizebox{\textwidth}{!}{
\footnotesize
\begin{tabularx}{\textwidth}{@{} l l l l l X @{}}
\toprule
\textbf{Dataset Name} & \textbf{Ref.} & \textbf{Year} & \textbf{Extraction Tool} & \textbf{Hardware} & \textbf{Specifications} \\ 
\midrule
UT-HAR (OP) & \cite{8067693} & 2017 & Linux 802.11n & Intel 5300 NIC & 3 $R_m$ Antennas, 30 Subcarriers Per Antenna, IEEE 802.11n/ac (5 GHz) \\ 
\midrule
WiCount (OP) & \cite{8367378} & 2017 & Linux 802.11n & Intel 5300 NIC, Mini R1C & 2 $T_n$, 3 $R_m$ Antennas, 30 Subcarriers, IEEE 802.11n (5 GHz, 40 MHz) \\ 
\midrule
CrossSense (OP) & \cite{10.1145/3241539.3241570} & 2018 & Linux 802.11n & Intel 5300 NIC, TP-Link WDR7500 & 3 $T_n$, 3 $R_m$ Antennas, 30 Subcarriers, IEEE 802.11n (5 GHz) \\ 
\midrule
WiADG (OP) & \cite{8487345} & 2018 & Atheros CSI & 2 TP-Link TL-WDR4300 & 3 $T_n$, 3 $R_m$ Antennas, 114 Subcarriers, IEEE 802.11n (5 GHz, 40 MHz) \\ 
\midrule
FallDeFi (OP) & \cite{10.1145/3161183} & 2018 & Linux 802.11n & Intel 5300 NIC & 2 $T_n$, 2 $R_m$ Antennas, 30 Subcarriers, IEEE 802.11n (5 GHz, 20 MHz) \\ 
\midrule
SignFi (OP) & \cite{10.1145/3191755} & 2018 & OpenRF 802.11n & Intel 5300 NIC & 3 $T_n$, 1 $R_m$ Antennas, 30 Subcarriers, IEEE 802.11n (5 GHz, 20 MHz) \\ 
\midrule
CSI Human Activity & \cite{app11198860} & 2019 & Nexmon CSI & Raspberry Pi 3B+/4B, ASUS RT-AC86U & 2 $T_n$, 2 $R_m$ Antennas, 128 Subcarriers (40 MHz), 256 Subcarriers (80 MHz) \\ 
\midrule
WiAR & \cite{8866726} & 2019 & Linux 802.11n & Intel 5300 NIC & 1 $T_n$, 3 $R_m$ Antennas, 30 Subcarriers, IEEE 802.11n (5 GHz, 20 MHz) \\ 
\midrule
Wi-SL & \cite{s20144025} & 2020 & Atheros CSI & TP-Link WDR4310 & 2 $T_n$, 3 $R_m$ Antennas, 114 Subcarriers, IEEE 802.11n (5 GHz, 40 MHz) \\ 
\midrule
DeepSeg (OP) & \cite{9235578} & 2020 & Linux 802.11n & Intel 5300 NIC & 1 $T_n$, 3 $R_m$ Antennas, 30 Subcarriers, IEEE 802.11n \\ 
\midrule
ZigFi (OP) & \cite{8970452} & 2020 & Linux 802.11n & Intel 5300 NIC & 3 $T_n$, 3 $R_m$ Antennas, 64 Subcarriers, IEEE 802.11n (2.4 GHz, 20 MHz) \\ 
\midrule
EyeFi (OP) & \cite{9183685} & 2020 & Linux 802.11n & Intel 5300 NIC & 3 $R_m$ Antennas, 30 Subcarriers, IEEE 802.11n \\ 
\midrule
Widar3.0 (OP) & \cite{7znf-qp86-20} & 2020 & Linux 802.11n & Intel 5300 NIC & 1 $T_n$, 18 $R_m$ Antennas, 30 Subcarriers, IEEE 802.11n (5.8 GHz) \\ 
\midrule
Wi-Fi CSI-based HAR & \cite{ZHURAVCHAK202259} & 2022 & Atheros CSI & 2 TP-Link TL-WDR4300 & 3 $T_n$, 3 $R_m$ Antennas, 56 Subcarriers (2.4 GHz, 20 MHz), 114 Subcarriers (5 GHz, 40 MHz) \\ 
\midrule
eHealth CSI & \cite{10177905} & 2023 & Nexmon CSI & Raspberry Pi 4B & 1 $T_n$, 1 $R_m$ Antenna, 234 Subcarriers, IEEE 802.11n (5 GHz, 80 MHz) \\ 
\midrule
IEEE 802.11ac 80 MHz & \cite{10144501} & 2023 & Nexmon CSI & Netgear X4S, ASUS RT-AC86U & 3 $T_n$, 3 $R_m$ Antennas, 242 Subcarriers, IEEE 802.11ac (5 GHz, 80 MHz) \\ 
\midrule
WiMANS (OP) & \cite{huang2024wimans} & 2024 & Linux 802.11n & Intel 5300 NIC & 3 $T_n$, 3 $R_m$ Antennas, 30 Subcarriers, IEEE 802.11n (2.4 GHz, 5 GHz) \\ 
\bottomrule
\end{tabularx}
}
\end{table*}

\section{Preprocessing Techniques for CSI Data} \label{sec:preprocessing}

Data preprocessing is a critical step in ML as it directly impacts the performance and generalization of the model. Real-world data such as CSI, often contains noise, missing values, outliers, and irrelevant or redundant features, which can lead to issues like overfitting bias, and poor deployment results. Preprocessing techniques such as cleaning, normalization, feature selection, and transformation address these challenges by reducing dimensionality, enhancing data quality, and ensuring compatibility with algorithms. These steps make learning algorithms more efficient and improve the accuracy and the interpretability of results \cite{kotsiantis2006data}.

Although DL algorithms can inherently learn the underlying patterns within CSI data, their performance is significantly enhanced when the data is preprocessed to highlight relevant features and reduce noise. Preprocessing also helps mitigate the inherent nature of CSI data, such as noise, multi-path effects, and interference, which can obscure the information of interest. The complex nature of CSI  often leads to signal distortions, making it challenging to focus on specific features, such as body movement in motion recognition applications. To overcome these challenges, a variety of preprocessing techniques have been developed. These techniques range from signal transformation methods that create new feature spaces to outlier removal and dimensionality reduction strategies. Each technique is tailored to enhance or suppress irrelevant features, depending on the specific application and data characteristics. The following subsections explore some of the most commonly used preprocessing methods and their practical applications in Wi-Fi sensing.

\subsection{Outlier Removal}

Outlier removal is essential in preprocessing signals like CSI and internal measurement (IMU)\footnote{An IMU is a sensor device that provides raw motion data such as linear acceleration, angular velocity, and orientation, typically via embedded accelerometers and gyroscopes.} data, which are often affected by noise and variability. Different techniques are used to identify and remove outliers as described below.

\subsubsection{Simple Smoothing Techniques}
Simple smoothing techniques reduce random noise by averaging data points within a set window. For example, the Moving Average (MA) method replaces each data point \(x_i\) with the average of neighbouring points in a window of size \(w\):
\begin{equation} 
\hat{x}_i = \frac{1}{w} \sum_{j=i-\frac{w-1}{2}}^{i+\frac{w-1}{2}} x_j
\end{equation}
The median filter further enhances robustness against outliers by substituting each data point with the median of its neighbours. However, these methods can sometimes overly smooth critical signal variations. To counter this, Weighted Moving Average (WMA) and Exponentially Weighted Moving Average (EWMA) apply weights \(w_j\) to emphasize recent data points as formulated below:
\begin{equation}
    \hat{x}_i = \sum_{j=i-\frac{w-1}{2}}^{i+\frac{w-1}{2}} w_j x_j \quad \text{where} \quad \sum_j w_j = 1.
\end{equation}
\subsubsection{Advanced Filtering Methods}
Advanced filtering methods focus on specific frequency components in the signal. Low-pass filters (LPF) reduce high-frequency noise by selecting an appropriate cutoff frequency \(f_c\), which helps preserve essential signal components. Wavelet filters, using the Discrete Wavelet Transform (DWT), break down the signal into various frequency bands, enabling precise noise reduction by retaining only selected wavelet coefficients.

\subsubsection{Anomaly Detection Techniques}
Anomaly detection methods identify and correct data points that deviate from expected patterns. The Hampel Filter flags outliers by comparing each data point \(x_i\) to the median \(m_i\) of neighbouring points and replacing points where \(|x_i - m_i|\) exceeds a threshold. Another method, the Local Outlier Factor (LOF) \cite{10.1145/3310194} assesses the local density around each point, marking outliers as those with substantially lower density compared to their neighbours. 

\subsection{Phase offset Compensation}

In practical Wi-Fi systems, various sources of noise arise from the transmitter and receiver processing stages, as well as hardware and software imperfections. These noise factors significantly affect the phase of the measured CSI between transmitting and receiving antennas, introducing complexities such as Cyclic Shift Diversity (CSD), Sampling Time Offset (STO), Sampling Frequency Offset (SFO), and beamforming-induced phase variations. Since Wi-Fi sensing applications rely on extracting the underlying multipath channel information to detect changes in the environment, addressing these phase distortions through appropriate signal processing techniques is crucial \cite{10734771}. A few important techniques are outlined below.

\subsubsection{Common Phase Error (CPE) Compensation}
In CPE compensation, it is assumed that each subcarrier’s phase is affected by a common phase error \(\theta_{\text{CPE}}\) in addition to the actual phase \(\theta_n\) of subcarrier \(n\):
\begin{equation}
\tilde{\theta}_n = \theta_n + \theta_{\text{CPE}}
\end{equation}
CPE compensation estimates \(\theta_{\text{CPE}}\) and subtracts it from the measured phase \(\tilde{\theta}_n\) across all subcarriers:
\begin{equation}
\hat{\theta}_n = \tilde{\theta}_n - \hat{\theta}_{\text{CPE}}
\end{equation}
\noindent where \(\hat{\theta}_n\) denotes the compensated phase, aligning phase data across subcarriers to improve accuracy in subsequent analysis.

\subsubsection{Phase Difference Method (PDM)}
 calculates phase differences between adjacent subcarriers or time samples. For adjacent subcarriers with phases \(\tilde{\theta}_n\) and \(\tilde{\theta}_{n+1}\), the phase difference \(\Delta \tilde{\theta}_{n,n+1}\) is given by:
\begin{equation}
\Delta \tilde{\theta}_{n,n+1} = \tilde{\theta}_{n+1} - \tilde{\theta}_n
\end{equation}
This method cancels out constant phase offsets, leaving stable phase differences \(\Delta \theta_{n,n+1}\) that are beneficial for applications like motion detection and device-free localization \cite{Abhayawardhana2002CPE,wagdy1987phasePDM}.

\subsubsection{CSI Ratio Model}
The CSI Ratio Model is another technique for mitigating common phase noise and compensating for phase offsets in Wi-Fi routers, particularly for those equipped with multiple antennas \cite{wu2022wifi}. In such systems, antennas are placed in close proximity, causing them to experience nearly identical phase shifts due to common noise sources like oscillator drift or environmental factors. By computing the ratios of the CSI measurements between pairs of antennas, the common phase components cancel out, isolating the relative phase differences caused by unique path effects to each antenna. Mathematically, the CSI ratio model is expressed as:
\begin{equation}
\frac{H_i(f)}{H_j(f)} = \frac{|H_i(f)| \, e^{j[\theta_i(f) + \phi_c(f)]}}{|H_j(f)| \, e^{j[\theta_j(f) + \phi_c(f)]}} = \frac{|H_i(f)|}{|H_j(f)|} \, e^{j[\theta_i(f) - \theta_j(f)]}
\end{equation}
In this equation, \( H_i(f) \) and \( H_j(f) \) are the CSI measurements at frequency \( f \) for antennas \( i \) and \( j \), respectively. The magnitudes of these measurements are \( |H_i(f)| \) and \( |H_j(f)| \), while \( \theta_i(f) \) and \( \theta_j(f) \) represent the unique phase shifts for each antenna. The term \( \phi_c(f) \) denotes the common phase noise experienced by all antennas, and \( j \) is the imaginary unit. By taking the ratio, the common phase term \( \phi_c(f) \) cancels out, allowing for better phase offset compensation based on the relative phase differences \( \theta_i(f) - \theta_j(f) \). Though this technique does not provide absolute phase values, they effectively capture relative phase changes, enhancing the utility of phase information for ML algorithms in Wi-Fi sensing.

\subsection{Dimension Reduction / Compression}

Dimension reduction is crucial for managing the high dimensionality of CSI data. It reduces noise by eliminating irrelevant features, simplifying models to be less prone to overfitting and more computationally efficient.

\subsubsection{Principal Component Analysis (PCA)}
PCA transforms high-dimensional data into a smaller set of orthogonal components while retaining as much variance as possible. Given a data matrix \(\mathbf{X} \in \mathbb{R}^{n \times p}\), PCA finds a projection matrix \(\mathbf{W} \in \mathbb{R}^{p \times k}\) to map \(\mathbf{X}\) into a lower-dimensional space:
\begin{equation}
\mathbf{Z} = \mathbf{X} \mathbf{W}
\end{equation}
\noindent where \(\mathbf{Z} \in \mathbb{R}^{n \times k}\) represents the reduced data. To improve data quality, it’s often useful to remove outliers before applying PCA, as it is sensitive to extreme values \cite{shi2022environment}.

\subsubsection{Low-Rank Factorization (LRF)}
Low-Rank Factorization approximates the data using lower-rank matrices. For a data matrix \(\mathbf{X} \in \mathbb{R}^{n \times p}\), LRF identifies matrices \(\mathbf{U} \in \mathbb{R}^{n \times r}\) and \(\mathbf{V} \in \mathbb{R}^{p \times r}\) such that:
\begin{equation}
\mathbf{X} \approx \mathbf{U} \mathbf{V}^\top
\end{equation}
This technique is useful for applications like matrix completion, offering greater flexibility than PCA for targeted dimensionality reduction. These dimension reduction techniques enhance CSI data processing, improving both model performance and computational efficiency.

\subsection{Signal Transform Methods}

Signal transforms  are fundamental for uncovering activity-related patterns or generating more informative representations suitable for training ML models. Key techniques include the Fast Fourier Transform (FFT), Short-Time Fourier Transform (STFT), and Wavelet Transform (WT) for the time-frequency analysis of CSI measurements. Additionally, specialized approaches such as Doppler Velocity Extraction and the Body-coordinate Velocity Profile (BVP) for estimating movement velocity and providing a robust representation of static environmental factors, as well as Random Convolutional Kernels for comprehensive feature extraction, are employed.

\subsubsection{Fast Fourier Transform (FFT)}
FFT is widely used to convert a time-domain signal \(x(t)\) into its frequency-domain representation \(X(f)\):
\begin{equation}
X(f) = \sum_{n=0}^{N-1} x(n) e^{-j2\pi fn/N}
\end{equation}
where \(N\) is the number of points, \(f\) is the frequency, and \(j\) is the imaginary unit. FFT reveals dominant frequencies in the signal and, when combined with Low Pass Filters (LPF) or Bandpass Filters (BPF), effectively removes high-frequency noise or targets specific frequency ranges, making it useful for applications like motion detection and breath estimation \cite{shi2022environment}.

\subsubsection{Short-Time Fourier Transform (STFT)}
STFT analyzes the frequency content of a signal over time by applying FFT to short, overlapping segments of the signal \(x(t)\) with a windowing function \(w(t)\):
\begin{equation}
\text{STFT}(t,f) = \sum_{n=-\infty}^{\infty} x(n)w(n-t)e^{-j2\pi fn}
\end{equation}
This method provides a time-frequency representation, useful for identifying patterns that vary over time. However, STFT encounters trade-offs between time and frequency resolution.

\subsubsection{Wavelet Transform (WT)}
The Wavelet Transform offers better frequency resolution for low frequencies and better time resolution for high frequencies. The Discrete Wavelet Transform (DWT) decomposes a signal into multiple resolution levels using high-pass and low-pass filters:
\begin{equation}
x(t) = \sum_{k=-\infty}^{\infty} c_k \phi_k(t) + \sum_{k=-\infty}^{\infty} \sum_{j=1}^{J} d_{j,k} \psi_{j,k}(t)
\end{equation}
\noindent where \(c_k\) are approximation coefficients, \(d_{j,k}\) are detail coefficients, \(\phi_k(t)\) is the scaling function, and \(\psi_{j,k}(t)\) are the wavelet functions at scale \(j\). DWT is effective for noise reduction and offers robustness beyond techniques relying solely on Doppler phase shift \cite{10.1145/2789168.2790093}.

\subsubsection{MUSIC}
CSI and the velocity of movement or activity are theoretically related, as shown in Eq.\ref{equation:12}. This relationship suggests that velocities of moving objects within the environment can be determined from CSI using AoA estimation methods, such as Multiple Signal Classification (MUSIC) \cite{schmidt1986multiple}. MUSIC leverages the eigen structure of the covariance matrix of the CSI data to estimate the movements Doppler velocity vector by separating the signal subspace from the noise subspace. The Doppler velocity pseudo-spectrum $P_{\text{MUSIC}}(v)$ is given by:
\begin{subequations}
\begin{align}
P_{\text{MUSIC}}(v) 
& = \frac{1}{\mathbf{d}^H(v) \mathbf{Q}_n \mathbf{Q}_n^H \mathbf{d}(v)}
\end{align}
\label{equation:18}
\end{subequations}
{where $\mathbf{d}$ is the steering vector for velocity $v$, and $\mathbf{Q}_n$ is the matrix of all eigenvectors forming the noise subspace. The denominator becomes zero when $v$ corresponds to the velocity of one of the moving points. Therefore, the sharp peaks in the pseudo-spectrum provide the estimated velocities of the moving points in the environment.}

\subsubsection{Body-coordinate Velocity Profile (BVP)}
BVP, introduced in systems like Widar3.0 \cite{9516988}, is designed for human motion analysis by leveraging unique velocity distributions across body parts during activities. A key parameter in BVP is the Doppler Frequency Shift (DFS), calculated as:
\begin{equation}
f_D = \frac{v \cdot f_c}{c}
\end{equation}
\noindent where \(f_D\) is the Doppler shift, \(v\) is relative velocity, \(f_c\) is the carrier frequency, and \(c\) is the speed of light. While DFS provides valuable insights, it is sensitive to position and orientation. BVP addresses this by mapping signal power over body-coordinated velocity components, making it a robust indicator of activity types.

\subsubsection{MiniRocket}

Due to the typically small size of datasets in HAR tasks, deep neural networks may struggle to effectively learn patterns from limited data. Unlike DL models that require large labeled datasets, MiniRocket, a highly effective method for time-series analysis, enables feature extraction without the need for labels, making it particularly advantageous in scenarios with scarce data \cite{dempster2021minirocket}.

In the MiniRocket method, convolution is applied to extract features from time-series data, such as CSI signals, by using random kernels. Each kernel 
$k$ shifts along the input signal 
$x$, generating an output 
$(x * k)_i$
, which captures localized patterns. These diverse transformed outputs are summarized using the Proportion of Positive Values (PPV), calculated as:
\begin{equation}
\operatorname{PPV}=\frac{1}{N} \sum_{i=1}^N \mathbb{I}\left((x * k)_i>0\right)
\end{equation}
where $N$ is the number of elements in the convolution output, and $\mathbb{I}(\cdot)$ is an indicator function that returns $1$ when the condition inside is true.  These extracted features, which in the default setting amount to $9,996$ features, can then be used as inputs to a simple linear classifier.

\section{Conventional DL  for Wi-Fi Sensing} 
\label{sec:dl}

In this section, we provide a brief overview of the existing DL methods for CSI-based Wi-Fi sensing. As summarized in \textbf{Table~\ref{table:method_selection}},  time-independent models (e.g., CNNs, DNNs) suited for spatial feature extraction and time-dependent models (e.g., RNNs, LSTMs) are suitable for capturing temporal dynamics. Sequential tasks benefit from models like RNNs, GRUs, and Transformers, which handle temporal dependencies effectively. In contrast, spatial tasks leverage CNNs for modeling spatial correlations across antennas and subcarriers. For scenarios with limited labeled data, unsupervised methods such as autoencoders and SSL approaches are invaluable for learning meaningful representations. 

\subsection{Time-Independent Models}
\label{sec:time-independent}
Wi-Fi sensing environments are highly dynamic and require models capable of adapting to temporal shifts. Time-series datasets are influenced by external factors such as movement and interference, making time-dependent models essential for accurately capturing temporal changes. However, time-independent models process data without considering temporal dependencies, making them well-suited for tasks where sequence order does not influence outcomes.

\paragraph{Deep Neural Networks (DNNs)}
\label{subsec:dnn}

DNNs \cite{sze2017efficient} model complex nonlinear interactions through multiple layers of connected neurons, enabling them to effectively extract patterns from high-dimensional CSI data. Each layer \( l \) computes an activation \( \mathbf{h}^{(l)} \) as follows:
\begin{equation}
\mathbf{h}^{(l)} = \sigma\left( \mathbf{W}^{(l)} \mathbf{h}^{(l-1)} + \mathbf{b}^{(l)} \right),
\end{equation}
where \( \mathbf{h}^{(0)} = \mathbf{x} \) (input layer) and \( \mathbf{h}^{(L)} = \mathbf{y} \) (output layer). By stacking layers, DNNs extract hierarchical features, reducing the reliance on manual feature engineering. 
WiCount \cite{liu2017wicount} employed a fully connected feed-forward neural network with two hidden layers trained on the phase and amplitude features of CSI data. This system achieved an accuracy of 82.3\% in estimating up to five individuals in diverse indoor environments. 

\paragraph{Convolutional Neural Networks (CNNs)}
\label{subsec:cnn}

CNNs \cite{gu2018recent} are powerful feature extractors for spatial patterns in CSI data. The convolution operation applies a filter \( \mathbf{w} \) across the input \( \mathbf{x} \), producing feature maps \( \mathbf{y} \):
\begin{equation}
\mathbf{y} = f(\mathbf{x} * \mathbf{w} + \mathbf{b}),
\end{equation}
where \( * \) denotes convolution, \( \mathbf{b} \) is a bias term, and \( f \) is an activation function. CNNs are particularly effective in Wi-Fi sensing tasks, where spatial dependencies in CSI are critical for understanding human activity. Their ability to extract spatial features makes them indispensable for such applications. Typically, CNNs serve as feature extractors, with their output passed to a linear classifier for final prediction.

In device-free health monitoring, \cite{kumar2022cnn} utilized CNNs as a preprocessing technique connected to an SVM classifier for health status prediction, achieving improved accuracy. 
Similarly, \cite{nadia2023cnn} demonstrated that using CNNs as a backbone for feature extraction in HAR achieved state-of-the-art accuracy of 97.14\% on the UCI HAR dataset. This outperformed traditional models such as DNNs, LSTMs, and CNN-LSTMs. The CNN-MLP hybrid approach showcased its ability to reduce computational overhead while achieving superior results in low-resource environments.

\subsection{Time-Dependent Models}  
\label{sec:time-dependent}

Time-dependent models analyze sequential CSI data to uncover temporal dependencies, which are essential for dynamic tasks such as HAR.

\paragraph{Recurrent Neural Networks (RNNs)}

RNNs \cite{salehinejad2017recentRNN} are designed to learn sequential patterns by maintaining internal states over time. The hidden state \( \mathbf{h}_t \) at time \( t \) is computed as:
\begin{equation}
\mathbf{h}_t = \sigma(\mathbf{W}_h \mathbf{h}_{t-1} + \mathbf{W}_x \mathbf{x}_t + \mathbf{b}_h),
\end{equation}
where \( \sigma \) is an activation function, \( \mathbf{W}_h \) and \( \mathbf{W}_x \) are weight matrices, and \( \mathbf{b}_h \) is a bias vector.
{
\cite{hoang2019recurrent} proposed a fingerprinting approach for Wi-Fi-based localization using RNNs, leveraging sequential RSSI patterns to achieve an average localization error of 0.75 m, outperforming KNN and probabilistic models by 30\%. Similarly, \cite{ding2020wifi} developed a passive fall detection system using Wi-Fi signals. By combining RNNs with discrete wavelet transforms for preprocessing, their system significantly enhanced elder healthcare by achieving higher accuracy than traditional ML models.}
However, standard RNNs struggle with vanishing gradients, limiting their ability to learn long-term dependencies \cite{9748867}.

\paragraph{Long Short-Term Memory (LSTM)}
LSTM networks \cite{sherstinsky2020fundamentals} are specifically designed to overcome the limitations of standard RNNs by introducing sophisticated memory mechanisms. The LSTM architecture uses three gates to control information flow:
\begin{align}
    \mathbf{f}_t &= \sigma(\mathbf{W}_f \mathbf{x}_t + \mathbf{U}_f \mathbf{h}_{t-1} + \mathbf{b}_f) \quad \text{(Forget gate)}, \nonumber \\
    \mathbf{i}_t &= \sigma(\mathbf{W}_i \mathbf{x}_t + \mathbf{U}_i \mathbf{h}_{t-1} + \mathbf{b}_i) \quad \text{(Input gate)}, 
\nonumber\\
    \mathbf{o}_t &= \sigma(\mathbf{W}_o \mathbf{x}_t + \mathbf{U}_o \mathbf{h}_{t-1} + \mathbf{b}_o) \quad \text{(Output gate)}, \nonumber\\
    \mathbf{C}_t &= \mathbf{f}_t \odot \mathbf{C}_{t-1} + \mathbf{i}_t \odot \tanh(\mathbf{W}_c \mathbf{x}_t + \mathbf{U}_c \mathbf{h}_{t-1} + \mathbf{b}_c) \, \text{(Cell state)}, \nonumber\\
    \mathbf{h}_t &= \mathbf{o}_t \odot \tanh(\mathbf{C}_t) \quad \text{(Hidden state)}.\nonumber
\end{align}
Recent applications highlight LSTM's versatility in Wi-Fi sensing. \cite{abuhoureyah2024multi} improved LSTM performance by preprocessing Wi-Fi signals with Independent Component Analysis (ICA) and Continuous Wavelet Transform (CWT), achieving 97\% accuracy in distinguishing multiple users' activities. For network security, \cite{tu2024lstm} developed an LSTM-based system for real-time jamming detection in IoT networks, achieving 99.5\% accuracy and outperforming traditional time-independent models. Furthermore, \cite{fu2024wi} introduced Wi-SensiNet, combining CNN feature extraction with Bidirectional LSTM (BiLSTM) \cite{huang2015bidirectional} to provide forward and backward temporal context. This hybrid architecture achieved 99\% accuracy in through-wall activity recognition, setting a new state-of-the-art performance.


\paragraph{Advanced Temporal Models}
Recent architectural advancements have further refined temporal modeling, enabling better capture of short- and long-term dependencies. {
TCD-FERN \cite{timeSelective2024RNN} introduced a dual-network approach where one network processes immediate spatial features and another captures movement patterns, with selective attention mechanisms determining the most relevant temporal information. This architecture achieved breakthrough performance in multi-room human presence detection by distinguishing between static and dynamic Wi-Fi signal patterns. Similarly, BiVTC \cite{zandi2024robofisense} adapted vision transformers to Wi-Fi sensing by treating CSI measurements as image-like data, enabling precise robotic activity recognition through attention mechanisms that identify subtle signal variations.}
These approaches build on the fundamental attention mechanism:
\begin{equation}
\text{Attention}(\mathbf{Q}, \mathbf{K}, \mathbf{V}) = \text{softmax}\left(\frac{\mathbf{Q} \mathbf{K}^\top}{\sqrt{d_k}}\right)\mathbf{V},
\end{equation}
where \( \mathbf{Q}, \mathbf{K}, \mathbf{V} \) are query, key, and value matrices. While attention-based models show immense promise, their computational demands pose challenges \cite{lin2022survey}, prompting the development of efficient alternatives like xLSTM \cite{beck2024xlstm}, which balance sophisticated temporal modeling with resource efficiency.

\subsection{Unsupervised Learning Approaches}
\label{sec:unsupervised}

Unsupervised learning is critical  when labeled data is limited or expensive to obtain. These approaches excel at tasks like anomaly detection, clustering, and feature extraction by identifying patterns in unlabeled data. Their versatility makes them well-suited for real-world deployments, requiring less preprocessing. Techniques such as clustering, autoencoders, and dimensionality reduction have significantly improved the efficiency of Wi-Fi sensing systems.

\subsubsection{Clustering}
Clustering algorithms group data points based on shared features, enabling automatic pattern discovery in CSI data without requiring labeled samples. Common applications include detecting anomalous activities, such as unusual human movements, and segmenting environmental conditions like office or home settings.
A notable method, \( k \)-means clustering, minimizes within-cluster variance:
\begin{equation}
\underset{\{\boldsymbol{\mu}_i\}_{i=1}^k}{\operatorname{arg\,min}} \sum_{i=1}^{k} \sum_{\mathbf{x} \in C_i} \|\mathbf{x} - \boldsymbol{\mu}_i\|^2,
\end{equation}
where \( \boldsymbol{\mu}_i \) represents the centroid of cluster \( C_i \). Similarity between data points \( \mathbf{x} \) and \( \mathbf{x}' \) is often measured using the Euclidean distance:
\begin{equation}
d(\mathbf{x}, \mathbf{x}') = \sqrt{\sum_{j=1}^{n} (x_j - x'_j)^2}.
\end{equation}
Clustering has been employed in applications such as noise identification \cite{tang2004noise} and human behavior analysis \cite{koh2020multiple}. In \cite{koh2020multiple}, clustering is applied across three dimensions—time, person, and location—to extract distinct patterns from CSI data. Clustering by time uncovers periodic trends, clustering by person segments users based on behavioral traits, and clustering by location identifies spatial usage patterns.

\subsubsection{Autoencoders}
\label{subsub:autoencoder}
Autoencoders learn compact representations by encoding input \( \mathbf{x} \) into a latent space \( \mathbf{z} \) and reconstructing it as \( \hat{\mathbf{x}} \):
\begin{align}
\mathbf{z} &= f_{\text{enc}}(\mathbf{x}), \\
\hat{\mathbf{x}} &= f_{\text{dec}}(\mathbf{z}).
\end{align}
These models effectively denoise, detect anomalies, and learn robust representations of environmental changes. In Wi-Fi sensing, autoencoders filter noise from CSI data, enhancing feature quality for downstream tasks. Advanced architectures, such as CNNs  and DNNs  are often integrated with autoencoders to process spatial and temporal patterns in CSI data.

Autoencoders are typically paired with methods like few-shot learning. SwinFi \cite{bian2024swinfi} incorporates an autoencoder architecture to compress and reconstruct CSI data, reducing computational and communication overhead. Similarly, AutoSen \cite{gao2024autosen} utilizes a cross-modal autoencoder to enhance Wi-Fi sensing by linking CSI amplitude and phase, eliminating noise, and extracting robust features. These approaches demonstrate the versatility of autoencoders in optimizing both raw data quality and downstream processing efficiency.




\begin{table*}[h!]
    \centering
    \caption{Recommended DL Methods for CSI-based Wi-Fi Sensing Tasks}
    \small
    \label{table:method_selection}
    \begin{tabular}{p{3cm} p{3cm} p{9cm}}
    \toprule
    \textbf{Task Type} &\textbf{ Data Characteristics} & \textbf{Recommended Methods} \\
    \midrule
    Temporal and Sequential Analysis &
    Sequential CSI Data &
    RNNs, LSTMs, and Transformers for capturing temporal patterns in CSI sequences \cite{abuhoureyah2024multi,tu2024lstm,fu2024wi,huang2015bidirectional,timeSelective2024RNN}; \newline
    Classification tasks that rely on temporal dependencies can also leverage these methods. \\
    \midrule
    Spatial Analysis & 
    CSI Amplitude Maps & 
    CNNs for extracting spatial features from CSI amplitude \newline
    Attention mechanisms to focus on relevant antennas or subcarriers \cite{kumar2022cnn,nadia2023cnn,spatialanal}; \newline
    Classification tasks with strong spatial dependencies can also use these methods. \\
    \midrule
    Generative Modeling & 
    Amplitude and Phase Distributions & 
    Autoencoders for learning representations and compression of CSI data; \newline
    Generative usage for data augmentation or anomaly detection \cite{gao2024autosen, bian2024swinfi, yang2023sensefi} \\
    \midrule
    Unsupervised Learning & 
    Unlabeled CSI Data & 
    Clustering methods to discover patterns in CSI amplitude and phase \cite{koh2020multiple}; \newline
    SSL for pretraining representations \cite{xu2023selfsupervisedlearningwificsibased} \\
    \midrule
    Few-Shot Learning & 
    Limited Labeled CSI Data & 
    Transfer learning from related Wi-Fi sensing tasks \cite{yin2022fewsense, zhao2024oneisenough} \newline
    Fine-tuning pre-trained models on small labeled datasets. \\
    \bottomrule
    \end{tabular}
\end{table*}

\begin{table*}[h]
    \centering
    \caption{Comparison of learning paradigms in Wi-Fi sensing}
    \small
    \begin{tabular}{p{4cm}p{3.5cm}p{3.5cm}p{4cm}}
        \toprule
        \textbf{Aspect} & \textbf{Self-Supervised} & \textbf{Supervised} & \textbf{Unsupervised} \\
        \midrule
        Label Requirement & None & High & None \\
        Feature Type & Discriminative features & Task-specific features & Cluster/reconstruction features \\
        Data Efficiency & High & Low & High \\
        Generalization & High & Moderate & Poor \\
        \bottomrule
    \end{tabular}
    \label{tab:comparison_ssl_sl_ul}
\end{table*}

\section{SSL for Wi-Fi Sensing}
\label{sec:SSL}
{In comparison to traditional supervised and unsupervised learning approaches, as summarized in \textbf{Table~\ref{tab:comparison_ssl_sl_ul}}, SSL offers various advantages. Unlike supervised learning, which requires large number of labels and tends to overfit task-specific patterns, SSL learns from unlabeled data, reducing dependency on extensive annotations. It is noteworthy that
online annotations for Wi-Fi sensing CSI data is exceptionally tedious, as CSI signals are not easily interpretable through visualization and require substantial domain expertise to accurately annotate. This challenge is compounded by the fact that human activities in Wi-Fi sensing often involve subtle signal variations that are difficult to distinguish without specialized knowledge, making SSL's ability to learn from unlabeled data particularly valuable in this domain. SSL achieves this by solving pretext tasks designed to simulate supervised objectives, enabling the model to learn discriminative and transferable features directly from raw CSI signals. These learned representations can subsequently be fine-tuned using a relatively small amount of labeled data, resulting in improved performance on downstream tasks while promoting better generalization and adaptability across different sensing environments \cite{technologies9010002, xu2023selfsupervisedlearningwificsibased}.
}

 {On the other hand, compared to unsupervised learning, which often produces generic or cluster-based representations, SSL provides discriminative feature embeddings that are more suitable for complex downstream tasks. }

{In addition, we would like to emphasize that SSL models, by virtue of their task-agnostic feature extraction, are inherently better suited for domain adaptation~\cite{Sfda}, enabling effective transfer across different environments. This generalization capability is particularly valuable in Wi-Fi sensing where signal characteristics vary significantly across different environments, devices, and user behavior.
}

\begin{figure*}[ht]
    \centering
    
    \begin{subfigure}{0.33\textwidth} 
        \centering
        \includegraphics[width=\textwidth]{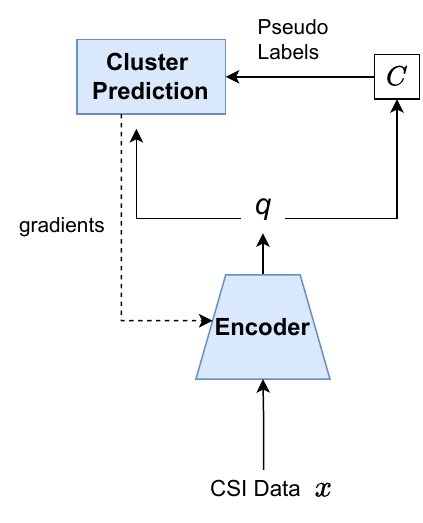}
        \caption{Cluster Discrimination Architecture}
        \label{subfig:ClustDisc}
    \end{subfigure}
    \hspace{0.02\textwidth} 
    \begin{subfigure}{0.29\textwidth} 
        \centering
        \includegraphics[width=\textwidth]{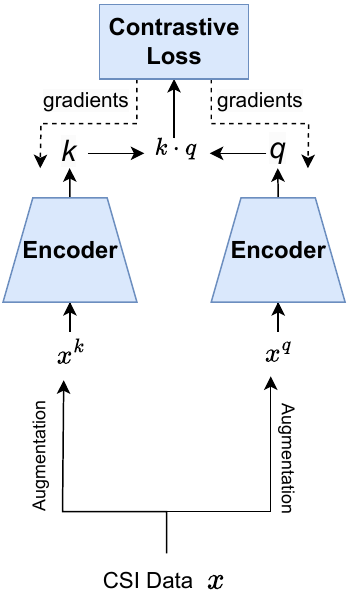}
        \caption{End-To-End}
        \label{subfig:EndToEnd}
    \end{subfigure}
    \hspace{0.02\textwidth} 
    \begin{subfigure}{0.29\textwidth} 
        \centering
        \includegraphics[width=\textwidth]{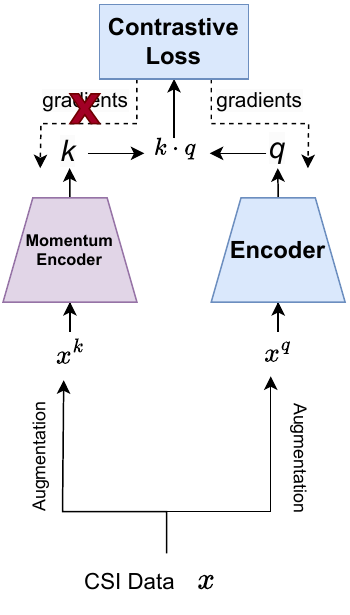}
        \caption{Momentum Encoder}
        \label{subfig:MomEnc}
    \end{subfigure}

    \vspace{2.5em} 

    \begin{subfigure}{0.48\textwidth} 
        \centering
        \includegraphics[width=\textwidth]{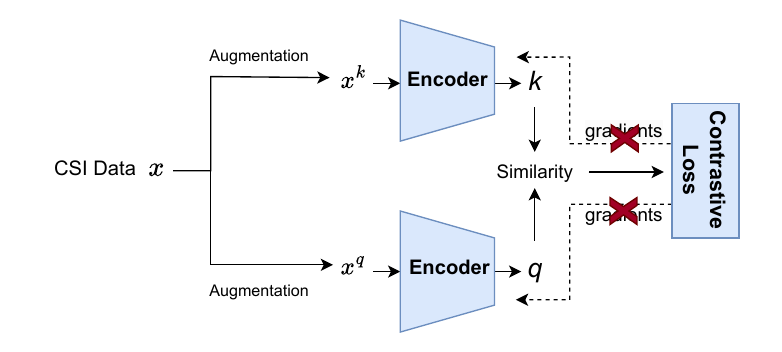}
        \caption{Instance Discrimination}
        \label{subfig:InsDisc}
    \end{subfigure}
    \hspace{0.01\textwidth} 
    \begin{subfigure}{0.49\textwidth} 
        \centering
        \includegraphics[width=\textwidth]{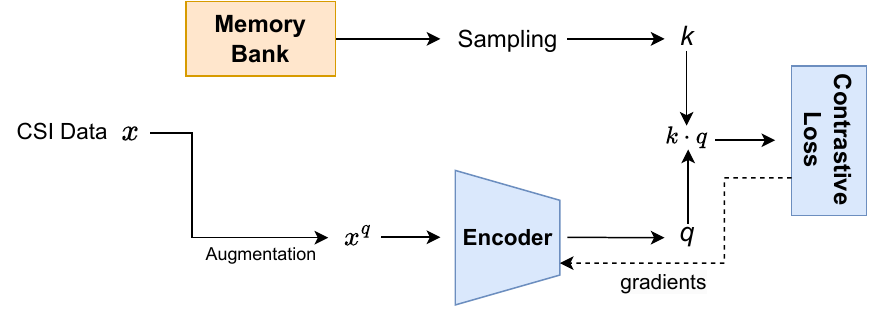}
        \caption{Memory Bank}
        \label{subfig:MemoryBank}
    \end{subfigure}
    \caption{Overview of SSL Architecture Paradigms}
    \label{fig:SSLArchParadigms}
\end{figure*}

\subsection{Theoretical Foundations of SSL for Wi-Fi CSI Sensing}
\label{subsec:SSL_theory}
{ Environmental dynamics, such as human motion and object displacement, introduce perturbations in CSI matrices—specifically in the amplitude \(|H_{m,n}(f_c,t)|\) and phase \(\angle H_{m,n}(f_c,t)\) components. These perturbations arise from variations in multi-path structure, Doppler shifts due to motion, and time-varying propagation delays. The amplitude component \(|H_{m,n}(f_c,t)|\) reflects large-scale variations due to reflection, scattering, and absorption, while the phase component \(\angle H_{m,n}(f_c,t)\) captures finer dynamics and is particularly sensitive to Doppler shifts. These frequency shifts encode motion information as a temporal pattern within the CSI phase.  Therefore, SSL can capture motion representations directly from phase evolution, even without explicit velocity or trajectory labels \cite{yang2022autofi}.}

{ In the context of SSL, two CSI measurements \( h^1_{f_c} \) and \( h^2_{f_c} \), corresponding to similar environmental states (e.g., the same activity or posture), are expected to differ only slightly. These observations can be expressed as follows:
\begin{equation}
h^2_{f_c} = h^1_{f_c} + \Delta h,
\end{equation}
where \(\Delta h\) captures small perturbations introduced by fine-grained movements such as breathing or posture shifts. SSL frameworks aim to learn an encoder \( f(\cdot) \) that maps such samples to a latent space where similar environmental conditions are placed closer together. This objective is commonly enforced by minimizing:
\begin{equation}
\mathcal{L}_{\text{SSL}} \propto \| f(h^1_{f_c}) - f(h^2_{f_c}) \|^2,
\end{equation}
with some methods also maximizing the distance between embeddings from different environmental conditions (i.e., negative pairs). }

\subsection{State-of-the-Art SSL Architectures}

SSL is structured around several core architecture paradigms, each contributing unique mechanisms for learning representations that are informative, robust, and transferable across domains. These paradigms, shown in \textbf{Fig.~\ref{fig:SSLArchParadigms}}, serve as the foundational building blocks of SSL and have inspired a diverse range of methods in the field. Rather than rigid frameworks, these paradigms are flexible components that can be combined in various ways, allowing SSL methods to incorporate multiple paradigms to address specific challenges.

\subsubsection{Cluster Discrimination Architectures}

Cluster discrimination extends traditional instance-based contrastive learning by grouping similar features into clusters that act as pseudo-labels. Instead of treating each sample independently, this approach allows models to learn features by associating samples within shared clusters. Swapping Assignments between Views (SwAV) \cite{caron2020unsupervisedSwav} exemplifies this paradigm by clustering embeddings and using these clusters as targets in a contrastive setting, effectively capturing high-level patterns while maintaining cohesive representations across batches \cite{NEURIPS2020_70feb62b}.

\subsubsection{End-to-End Contrastive Architectures}

End-to-end contrastive architectures operate as fully differentiable systems designed to optimize all components through gradient-based learning. They typically consist of two core elements: a Query encoder ($Q$) and a Key encoder ($K$). The Query encoder processes the original samples, while the Key encoder processes augmented views. A contrastive loss brings positive pairs closer in the embedding space while pushing negatives apart. This paradigm forms the basis of methods like the simple framework for contrastive learning of visual representations (SimCLR) \cite{chen2020simpleCLR}, momentum contrast (MoCo) \cite{he2020momentumMoCo}, and bootstrap your own latent (BYOL) \cite{grill2020byol}.

\subsubsection{Memory Bank Architectures}

Memory bank architectures address the need for large batch sizes by maintaining an external dictionary that stores feature representations across batches \cite{Wu_2018_CVPR}. This memory bank provides a diverse pool of negative samples, enhancing robustness and generalization without requiring all samples in the current batch. Representations within the memory bank are periodically updated, often through an exponential moving average, allowing models to decouple batch size from the number of available negatives. 

\subsubsection{Momentum Encoder Architectures}

Momentum encoder architectures improve representation learning efficiency by dynamically updating a queue of representations during training. This paradigm employs a momentum encoder that is gradually updated based on the parameters of the main encoder, stabilizing the learning process across mini-batches. The queue acts as a rolling dictionary, where new mini-batches are added, and the oldest are removed, ensuring a consistent set of embeddings. MoCo and BYOL exemplify this architecture, utilizing it to maintain temporally consistent features over training.

\subsubsection{Instance Discrimination Architectures}

Instance discrimination architectures focus on learning distinct representations for each sample by minimizing the distance between different augmented views of the same instance. Through augmentations such as cropping, flipping, and colour jittering, these architectures teach models to associate variations with the original sample, reinforcing instance-level distinctions in the embedding space. This paradigm has been foundational in SSL, enabling models to capture robust and invariant features without requiring predefined labels \cite{He_2020_CVPR}.

\subsubsection{Hybrid Architectures}

Many SSL methods incorporate elements from multiple paradigms, creating hybrid architectures that leverage the strengths of each approach. For instance, MoCo combines elements of momentum encoding with end-to-end contrastive learning, managing a dynamic queue for efficient negative sampling. BYOL integrates momentum encoding with instance discrimination, aligning augmented views without requiring negative pairs. SwAV merges cluster discrimination with instance discrimination, and clustering representations while ensuring alignment across views. This flexibility in combining paradigms allows SSL methods to address specific representation learning challenges, enhancing stability, scalability, and efficiency.

\subsection{Contrastive Learning Methods}
SSL approaches can be broadly categorized into contrastive and non-contrastive methods, each employing distinct strategies for representation learning. 

Contrastive learning in SSL trains an encoder to distinguish between data samples by maximizing similarity for positive pairs (similar samples) and minimizing it for negative pairs (dissimilar samples). This approach uses a contrastive loss function, like Noise Contrastive Estimation (NCE) \cite{gutmann2010noise}, to bring similar samples closer in representation space and push dissimilar samples further apart.

\subsubsection{SimCLR}
SimCLR is a contrastive SSL algorithm that learns meaningful representations by maximizing agreement between different augmentations of the same sample. As shown in \textbf{Fig.~\ref{subfig:InsDisc}}, SimCLR follows the instance discrimination paradigm, where an input CSI data sample $x$ is augmented to create two views $x^k$ and $x^q$. Each view is processed by the same encoder network, producing representations $k$ and $q$ respectively.

The objective of SimCLR is to maximize the similarity between the representations of the augmented pair while minimizing the similarity with other samples in the batch, known as negatives. This is achieved through a contrastive loss function called normalized temperature-scaled cross-entropy loss (NT-Xent). The loss is defined as:
\begin{equation}
\ell = -\log \frac{\exp(\text{sim}(k, q) / \tau)}{\sum_{i=1}^{2N} \mathbb{1}_{[i \neq k]} \exp(\text{sim}(k, k_i) / \tau)},
\end{equation}
where $\text{sim}(k, q)$ denotes the cosine similarity between $k$ and \(q\), defined as $\text{sim}(k, q) = \frac{k \cdot q}{\|k\| \|q\|}$, and $\tau$ is a temperature parameter that scales the similarities. SimCLR's simplicity and effectiveness, however, come with a computational cost, as it requires large batch sizes to generate a sufficient number of negative samples in each iteration. Despite this, SimCLR learns better representations \cite{chen2020simpleCLR}.

\subsubsection{MoCo (Momentum Contrast)}
MoCo enhances contrastive learning by framing it as a dictionary look-up task, designed to maintain a large and consistent set of negative samples without requiring large batch sizes. As shown in \textbf{Fig.~\ref{subfig:MemoryBank}} and \textbf{Fig.~\ref{subfig:MomEnc}}, MoCo combines two paradigms: a momentum encoder architecture and a memory bank to achieve scalable and stable representation learning.

For an input CSI data sample \(x\), two augmented views \(x^k\) and \(x^q\) are created. The query view \(x^q\) is encoded by the encoder to produce representation \(k\). For key views, a separate momentum encoder processes \(x^k\) to produce representation \(k \cdot q\). The parameters of the momentum encoder are updated more gradually than the query encoder, maintaining consistency by using a moving average of the query encoder's parameters.

The memory bank in MoCo acts as a queue, storing a large set of negative samples. This queue is continuously updated by enqueuing the newest keys and dequeuing the oldest ones, allowing MoCo to maintain a diverse set of negatives across batches without relying on large batch sizes. As shown in \textbf{Fig.~\ref{subfig:MemoryBank}}, the size of this memory bank is independent of the batch size, enhancing scalability.

The contrastive loss function aligns the query representation \(k\) with its corresponding key \(k \cdot q\) while pushing apart from other keys in the queue:
\begin{equation}
\ell = -\log \frac{\exp(\text{sim}(k, k \cdot q) / \tau)}{\sum_{i=0}^{K} \exp(\text{sim}(k, k_i) / \tau)},
\end{equation}
where \( \text{sim}(k, k \cdot q) \) is the cosine similarity between the representations, and \( \tau \) is a temperature parameter that scales the similarity. The use of a large queue size \( K \) allows for a substantial pool of negatives without increasing batch size.

The momentum encoder is updated as follows:
\begin{equation}
\theta_m \leftarrow m \theta_m + (1 - m) \theta,
\end{equation}
where \( \theta_m \) represents the parameters of the momentum encoder, \( \theta \) represents the parameters of the encoder, and \( m \) is the momentum coefficient. This update rule ensures smooth, consistent updates to the momentum encoder, preserving stable representations over time. By combining a momentum encoder with a memory bank, MoCo decouples dictionary size from batch size, achieving efficient and scalable contrastive learning. This design enables MoCo to maintain high representation quality without the computations of large batch sizes \cite{he2020momentumMoCo}.

\subsection{Non-Contrastive Learning Methods}

Non-contrastive SSL offers an alternative to contrastive approaches by learning representations without negative samples. This simplifies training and reduces memory usage, though it requires techniques to avoid the collapse problem, where all embeddings converge to a single point. Non-contrastive methods typically use architectural innovations or regularization methods to maintain diverse and informative representations.

\subsubsection{Bootstrap Your Own Latent (BYOL)}
 eliminates the need for negative samples by employing two networks: an online network and a target network. As in \textbf{Fig.\ref{subfig:MomEnc}}, BYOL's online network learns by predicting the representation generated by the target network from different augmented views of the same CSI data, effectively bootstrapping its latent representations.

For an input CSI sample, augmentations produce two views, denoted $v$  and $v'$, which are processed by the online encoder $f_\theta$ and the target encoder $f_\xi$, resulting in representations $y = f_\theta(v)$  and $y' = f_\xi(v')$. The online network further processes these through projection MLPs $g_\theta$ and $g_\xi$ to obtain $z = g_\theta(y)$ and $z' = g_\xi(y')$. Finally, a predictor MLP $q_\theta$ in the online network produces $q_\theta(z)$, which is trained to match $z'$ from the target network using a mean squared error (MSE) loss:
\begin{equation}
\mathcal{L} = \left\lVert \text{sg}(z') - q_\theta(z) \right\rVert^2_2,
\end{equation}
where $\text{sg}$ denotes the stop-gradient operation, preventing gradient updates from flowing back to the target network. This stop-gradient mechanism stabilizes the learning process, preventing the network from collapsing into trivial solutions where representations become identical or lose diversity.

To maintain stability and consistency in the target network, its parameters are updated via a slow-moving average of the online network parameters, defined as:
\begin{equation}
\xi \leftarrow m \xi + (1 - m) \theta,
\end{equation}
where $m$ is the momentum coefficient, $\xi$ denotes the target network parameters, and $\theta$ shows the online network parameters. By updating the target network gradually, BYOL ensures that the representations remain stable over time, thus preventing collapse where all outputs converge to a single point.

BYOL achieves competitive performance on datasets such as CSI data without relying on negative pairs, showing that meaningful self-supervised representations can be learned through positive pairs alone \cite{grill2020byol}. This paradigm shift has broadened the scope of SSL, demonstrating the effectiveness of non-contrastive approaches in representation learning.

\subsubsection{Swapping Assignments between Views (SwAV)}
SwAV is a clustering-based SSL method that enforces consistency between different views of the same CSI data, improving the quality of learned representations. As shown in \textbf{Fig.~\ref{subfig:ClustDisc}}, SwAV leverages cluster discrimination by assigning representations to prototype vectors C in a shared feature space.

Given an input CSI sample $x$, two augmented views $x^k$ and $x^q$ are generated through different transformations. These views are encoded into representations $q$ through the encoder network. SwAV then computes soft cluster assignments using a set of learnable prototype vectors $C$, which represent clusters in the feature space. The key innovation in SwAV is to optimize a swapped prediction task, where the cluster assignment of one view must be predicted from the representation of the other view.
The optimization process involves two main steps. First, computing soft cluster assignments $Q(q)$ by comparing the normalized embeddings with the prototype vectors $C$:
\begin{equation}
Q(q) = \text{softmax}\left(\frac{q^T C}{\tau}\right),
\end{equation}
where $\tau$ is a temperature parameter controlling the softness of the assignments.
Second, computing the cross-entropy loss between the cluster assignments of different views:
\begin{equation}
\mathcal{L} = \ell(q_1, Q(q_2)) + \ell(q_2, Q(q_1)),
\end{equation}
where
\begin{equation}
\ell(q, Q) = -\sum_{c} Q_c \log P_c(q),
\end{equation}
and $P(q)$ represents the predicted probability distribution over clusters for representation q.

To prevent trivial solutions, SwAV uses several  constraints. The prototype vectors $C$ are normalized to lie on the unit sphere. The assignments are balanced across the mini-batch to avoid cluster collapse. Additionally, an optimal transport solver is used to ensure uniform cluster assignment distributions. The optimization objective includes these constraints:
\begin{equation}
\min_{C, \theta} \mathbb{E}_{x \sim \mathcal{D}} [\mathcal{L}(x)] \quad \text{s.t.} \quad Q^T \mathbf{1} = \frac{B}{K} \mathbf{1},
\end{equation}
where $B$ is the batch size and $K$ is the number of prototypes.

SwAV has shown high accuracy on various datasets by combining the benefits of clustering and contrastive learning approaches. Its cluster discrimination approach provides several advantages. It requires no large memory banks or queue mechanisms. It naturally handles multiple views of the same sample. The method learns semantically meaningful cluster assignments while preserving instance-level information. Furthermore, it remains computationally efficient due to the absence of pairwise comparisons. The effectiveness of SwAV on CSI data opens new possibilities for SSL in Wi-Fi sensing applications, particularly where the underlying data exhibits natural clustering behaviour.

\subsubsection{Barlow Twins}
Barlow Twins \cite{zbontar2021barlow} focuses on redundancy reduction to ensure that representations of different views are both similar and uncorrelated. As shown in \textbf{Fig.\ref{subfig:InsDisc}}, Barlow Twins employs twin networks to process different augmented views of the same CSI data, aiming to make the cross-correlation matrix of their embeddings as close to an identity matrix as possible. This objective promotes invariance while minimizing redundancy across feature dimensions.

Given an input CSI sample $x$, two augmented views $x^k$ and $x^q$ are generated and processed by identical networks with shared parameters. The networks consist of an encoder $f(\cdot)$ followed by a projector network$ g(\cdot)$, producing embeddings $z^A = g(f(x^k))$ and $z^B = g(f(x^q))$. These embeddings are then used to compute the cross-correlation matrix between the normalized feature vectors along the batch dimension:
\begin{equation}
C_{ij} = \frac{\sum_b z^A_{b,i} z^B_{b,j}}{\sqrt{\sum_b (z^A_{b,i})^2} \sqrt{\sum_b (z^B_{b,j})^2}},
\end{equation}
where $b$ indexes batch samples, and $i$,$j$ index the feature dimensions. The objective function for Barlow Twins combines an invariance term and a redundancy reduction term:
\begin{equation}
\mathcal{L}_{BT} = \underbrace{\sum_{i}(1 - C_{ii})^2}_\text{invariance term} + \lambda \underbrace{\sum_{i}\sum_{j \neq i} C_{ij}^2}_\text{redundancy reduction term},
\end{equation}

The invariance term pushes the diagonal elements of $C$ toward 1, ensuring that corresponding features are maximally correlated. The redundancy reduction term pushes the off-diagonal elements of $C$ toward 0, ensuring that different features capture distinct information. The hyperparameter $\lambda$ controls the trade-off between these competing objectives.

This approach offers several unique advantages. The method naturally avoids collapsed solutions without requiring techniques like stop-gradient operations or momentum encoders. The cross-correlation computation implicitly takes into account the entire batch of samples, making it robust without requiring explicit negative samples or large batch sizes. Additionally, the redundancy reduction term helps learn more efficient representations by encouraging features to capture different aspects of the data.

The theoretical foundation of Barlow Twins draws inspiration from the Information Bottleneck principle and the redundancy reduction principle in neuroscience. By optimizing both invariance and decorrelation, the method learns representations that are both robust and efficient. The cross-correlation objective can be seen as approximating mutual information between views while encouraging independence between different feature dimensions.

When applied to CSI data, Barlow Twins can potentially capture important invariances while ensuring that different feature dimensions represent distinct aspects of the signal. The method's ability to work with reasonable batch sizes and its principled approach to redundancy reduction make it particularly suitable for wireless sensing applications. Furthermore, its straightforward implementation and theoretical grounding make it an attractive choice for scenarios where interpretable feature learning is desired.

\subsubsection{Variance-Invariance-Covariance Regularization (VICReg)}
VICReg \cite{bardes2021vicreg} is a SSL method that prevents collapse by enforcing three main properties in the embeddings: variance, invariance, and decorrelation. As shown in \textbf{Fig.~\ref{subfig:InsDisc}}, VICReg processes two augmented views of the input through the same encoder, but explicitly regularizes the structure of the embedding space through its unique loss function.

Given an input CSI sample $x$, two augmented views $x^k$ and $x^q$ are generated and encoded into embeddings $z^k$ and $z^q$. The VICReg loss function combines three terms, each addressing a specific aspect of representation learning:
\begin{equation}
\mathcal{L}_{\text{VICReg}} = \underbrace{\mathcal{L}_{\text{inv}}}_{\text{invariance}} + \lambda \underbrace{\mathcal{L}_{\text{var}}}_{\text{variance}} + \mu \underbrace{\mathcal{L}_{\text{cov}}}_{\text{covariance}},
\end{equation}
The invariance term $\mathcal{L}_{\text{inv}}$ ensures similarity between representations of different views:
\begin{equation}
\mathcal{L}_{\text{inv}} = \|z^k - z^q\|_2^2,
\end{equation}

The variance term $\mathcal{L}_{\text{var}}$ prevents dimensional collapse by ensuring each dimension of the representations has sufficient variance:
\begin{equation}
\mathcal{L}_{\text{var}} = \sum_i \max(0, \gamma - \text{Var}(z_i)),
\end{equation}
where \(\gamma\) is a target variance parameter.

The covariance term $\mathcal{L}_{\text{cov}}$ encourages decorrelation between different dimensions:
\begin{equation}
\mathcal{L}_{\text{cov}} = \sum_{i \neq j} \text{Cov}(z_i, z_j)^2,
\end{equation}
VICReg's approach offers several advantages. It explicitly controls representation structure through direct regularization rather than architectural constraints. The method works without batch normalization, momentum encoders, or predictor networks. Also, the explicit variance and covariance terms provide interpretable control over the learned representations.

\subsubsection{Simple Siamese (SimSiam)}
SimSiam \cite{Chen_2021_CVPR} demonstrates how simple architectural choices can prevent representation collapse without requiring negative samples or momentum encoders. As shown in \textbf{Fig.~\ref{subfig:InsDisc}}, SimSiam employs a siamese network architecture with an asymmetric component: a stop-gradient operation in one branch.

Given an input CSI sample $x$, two augmented views $x^k$ and $x^q$ are processed through an encoder network $f$ followed by a projection MLP $g$ and a prediction MLP $h$. The key asymmetry comes from applying a stop-gradient to one branch:
\begin{equation}
z_k = g(f(x^k)), \quad z_q = \text{sg}(g(f(x^q))),
\end{equation}
\begin{equation}
p_k = h(z_k),
\end{equation}
The loss function goal in SimSiam is to maximize the similarity between the siamese network output:
\begin{equation}
\mathcal{L} = \frac{1}{2} \left[ \mathcal{D}(p_k, z_q) + \mathcal{D}(p_q, z_k) \right],
\end{equation}
where $ \mathcal{D}(p, z) = - \frac{p}{\|p\|_2} \cdot \frac{z}{\|z\|_2}$  computes the negative cosine similarity. The stop-gradient operation is crucial in preventing collapse by creating an information bottleneck. It forces the network to learn meaningful representations that can predict the other view's features while maintaining diversity through the asymmetric structure. This simple yet effective design achieves competitive performance without the complexity of negative samples, large batches, or momentum encoders. The effectiveness of SimSiam demonstrates that the essential ingredients for successful SSL can be distilled into remarkably simple components. Its application to CSI data could offer an efficient approach to learning wireless signal representations, particularly in scenarios where computational resources are limited or when the training dynamics need to be well-understood and controlled.

\subsubsection{Masked Autoencoders (MAE)}
{Masked Autoencoders (MAE) \cite{he2022masked} offer a distinct approach to impute missing parts of the CSI signal compared to traditional SSL techniques by leveraging an autoencoder architecture as discussed in \textbf{Section~\ref{subsub:autoencoder}}. The input CSI data $x^k$ is partially masked and then passed to an asymmetric encoder that processes only the visible patches, mapping them into a latent representation $k$. A lightweight decoder subsequently takes $k$ along with mask tokens to reconstruct the original input, producing $\hat{x^k}$. The reconstruction loss is computed using the mean squared error (MSE) as follows:
\begin{equation}
    \mathcal{L}_{\text{MAE}} = \| x^k - \hat{x^k} \|^2_2
\end{equation}
Unlike contrastive methods such as SimCLR, MAE does not require dual encoder branches, leading to a more lightweight and computationally efficient design. However, due to its reconstruction-based objective, MAE typically requires longer training schedules to achieve optimal performance. This is because reconstructing fine-grained input details from sparsely observed patches is inherently a more challenging task than contrasting positive and negative pairs. Additionally, since the encoder processes only a subset of the input, it sees less information per iteration, necessitating more epochs for the model to fully capture the underlying structure of the data. Despite these longer training times, MAE yields highly transferable and semantically rich representations.}
{\subsubsection{Simple Framework for Masked Image Modeling (SimMIM)}
\cite{xie2022simmim} builds upon the concept of masked signal modeling by further simplifying the design introduced in MAE. In the context of Wi-Fi sensing, SimMIM applies random masking directly on the CSI input $x^k$, replacing selected patches with a learnable mask token, and passes the entire masked input through the encoder. Unlike MAE, which processes only the visible portions of the input, SimMIM retains the full structure, including masked areas. The encoder output is then fed into a lightweight linear prediction head that estimates the missing CSI values. The reconstruction is guided by an $\ell_1$ regression loss computed only over the masked patches:
\begin{equation}
    \mathcal{L}_{\text{SimMIM}} = \| x_M^k - \hat{x}_M^k \|_1
\end{equation}
where $x_M^k$ and $\hat{x}_M^k$ denote the ground truth and reconstructed masked CSI signals, respectively.
By eliminating the need for a complex decoder, SimMIM offers a highly efficient and scalable pretext task while achieving comparable fine-tuning performance to MAE. However, due to its focus on low-level signal recovery rather than semantic feature abstraction, SimMIM representations generalize less effectively without full fine-tuning. This limitation necessitates end-to-end adaptation during downstream Wi-Fi sensing tasks, in contrast to methods where only the classification head is fine-tuned.}

\subsection{Summary}

Contrastive methods such as SimCLR and MoCo leverage negative samples to learn discriminative representations. SimCLR relies on large batch sizes to provide sufficient negatives, while MoCo introduces a momentum encoder and a queue mechanism to maintain a dynamic set of negatives efficiently. These approaches effectively create well-separated embedding spaces but require careful negative sampling strategies. Non-contrastive methods, including BYOL, SwAV, Barlow Twins, VICReg, {MAE, and SimMIM}, eliminate the need for negative pairs through alternative mechanisms. BYOL employs asymmetric networks with a momentum-updated target network, while SwAV utilizes online clustering with prototype vectors. Barlow Twins and VICReg directly regularize embedding properties—Barlow Twins by reducing redundancy, and VICReg by enforcing variance, invariance, and covariance constraints. {Meanwhile, masked signal modeling approaches such as MAE and SimMIM learn representations by reconstructing missing parts of the input signal, offering an alternative strategy to avoid the reliance on contrastive pairs while improving scalability and pretraining efficiency.} These methods achieve competitive performance while reducing the complexity associated with contrastive sampling.

The key distinction lies in how representation collapse is prevented. Contrastive methods explicitly push negative pairs apart, while non-contrastive methods, including masked modeling, rely on architectural designs, reconstruction tasks, or direct feature regularization. Both paradigms have proven effective, with contrastive methods excelling when natural negatives are available, and non-contrastive methods offering greater flexibility and efficiency, summarized at \textbf{Table~\ref{SSLmodel_table}}. Future research explore combining these complementary strengths to develop more robust and versatile SSL frameworks.


\begin{table*}[h!]
\centering
\caption{A qualitative comparison of existing SSL Algorithms for Wi-Fi Sensing}
\small
\label{SSLmodel_table}
\resizebox{\textwidth}{!}{
\begin{tabularx}{\textwidth}{M{1.2cm} M{1.6cm} M{1.4cm} M{4.4cm} M{3.8cm} M{3.2cm}}
\toprule
\textbf{Name} & \textbf{Ref.} & \textbf{SSL Type} & \textbf{Advantages} & \textbf{Disadvantages} & \textbf{Notes} \\ 
\midrule
SimCLR & \cite{pmlr-v119-chen20j,9600010,10008537,9522151,10446566,liu2022self} & Contrastive & 
\begin{itemize}[nosep,leftmargin=*]
   \item Simple architecture without specialized components
   \item State-of-the-art performance with sufficient resources
\end{itemize} &
\begin{itemize}[nosep,leftmargin=*]
   \item Requires large batch sizes
   \item High computational demands
\end{itemize} &
\begin{itemize}[nosep,leftmargin=*]
   \item NT-Xent loss for view agreement
   \item Strong data augmentation dependency
\end{itemize} \\ 
\midrule
MoCo & \cite{He_2020_CVPR,xu2025evaluating} & Contrastive &  
\begin{itemize}[nosep,leftmargin=*]
   \item Dictionary independent of batch size
   \item Efficient negative sampling
\end{itemize} &
\begin{itemize}[nosep,leftmargin=*]
   \item Complex momentum update mechanism
   \item Careful queue management needed
\end{itemize} &
\begin{itemize}[nosep,leftmargin=*]
   \item Momentum encoder with queue
   \item InfoNCE loss
\end{itemize} \\ 
\midrule
SwAV & \cite{caron2020unsupervisedSwav,xu2025evaluating} & Non-Contrastive &  
\begin{itemize}[nosep,leftmargin=*]
   \item No negative pairs needed
   \item Efficient with small batches
\end{itemize} &
\begin{itemize}[nosep,leftmargin=*]
   \item Complex clustering mechanism
   \item Sensitive to prototype parameters
\end{itemize} &
\begin{itemize}[nosep,leftmargin=*]
   \item Online clustering approach
   \item Swapped prediction mechanism
\end{itemize} \\ 
\midrule
BYOL & \cite{grill2020byol} & Non-Contrastive &  
\begin{itemize}[nosep,leftmargin=*]
   \item No negative samples needed
   \item Robust to augmentation choices
\end{itemize} &
\begin{itemize}[nosep,leftmargin=*]
   \item Complex dual network setup
   \item Sensitive to momentum updates
\end{itemize} &
\begin{itemize}[nosep,leftmargin=*]
   \item Momentum encoder prevents collapse
   \item MSE prediction loss
\end{itemize} \\ 
\midrule
Barlow Twins & \cite{zbontar2021barlow,barahimi2024context} & Non-Contrastive &
\begin{itemize}[nosep,leftmargin=*]
   \item Simple siamese architecture
   \item Direct redundancy reduction
\end{itemize} &
\begin{itemize}[nosep,leftmargin=*]
   \item Memory-intensive correlation
   \item Careful loss balancing needed
\end{itemize} &
\begin{itemize}[nosep,leftmargin=*]
   \item Cross-correlation based
   \item Feature decorrelation objective
\end{itemize} \\ 
\midrule
VICReg & \cite{bardes2021vicreg} & Non-Contrastive &
\begin{itemize}[nosep,leftmargin=*]
   \item Explicit collapse prevention
   \item No architectural constraints
\end{itemize} &
\begin{itemize}[nosep,leftmargin=*]
   \item Complex three-term loss
   \item Sensitive to parameters
\end{itemize} &
\begin{itemize}[nosep,leftmargin=*]
   \item Variance-Invariance-Covariance loss
   \item Direct embedding regularization
\end{itemize} \\ 
\midrule
SimSiam & \cite{Chen_2021_CVPR} & Non-Contrastive &
\begin{itemize}[nosep,leftmargin=*]
   \item Extremely simple design
   \item Resource efficient
\end{itemize} &
\begin{itemize}[nosep,leftmargin=*]
   \item Sensitive to implementation
   \item Critical architecture choices
\end{itemize} &
\begin{itemize}[nosep,leftmargin=*]
   \item Stop-gradient prevents collapse
   \item Symmetric prediction loss
\end{itemize} \\ 
\midrule
MAE & \cite{he2022masked,10678770,10460073,xu2025evaluating} & Non-Contrastive &  
\begin{itemize}[nosep,leftmargin=*]
   \item Lightweight asymmetric encoder-decoder design
   \item Highly scalable for large models
\end{itemize} &
\begin{itemize}[nosep,leftmargin=*]
   \item Requires long training schedules
   \item Lower linear probing performance compared to contrastive methods
\end{itemize} &
\begin{itemize}[nosep,leftmargin=*]
   \item Reconstructs masked regions using MSE loss
   \item Strong fine-tuning results on downstream tasks
\end{itemize} \\ 
\midrule
SimMIM & \cite{xie2022simmim} & Non-Contrastive &  
\begin{itemize}[nosep,leftmargin=*]
   \item Extremely simple and efficient design
   \item Comparable fine-tuning performance to MAE
\end{itemize} &
\begin{itemize}[nosep,leftmargin=*]
   \item Focuses on low-level signal recovery
   \item Requires full fine-tuning for effective downstream adaptation
\end{itemize} &
\begin{itemize}[nosep,leftmargin=*]
   \item Predicts masked raw point with a linear head
   \item $\ell_1$ loss computed only on masked patches
\end{itemize} \\
\bottomrule
\end{tabularx}
}
\end{table*}

\begin{figure*}[ht]
    \centering
    \includegraphics[width=1.0\linewidth]{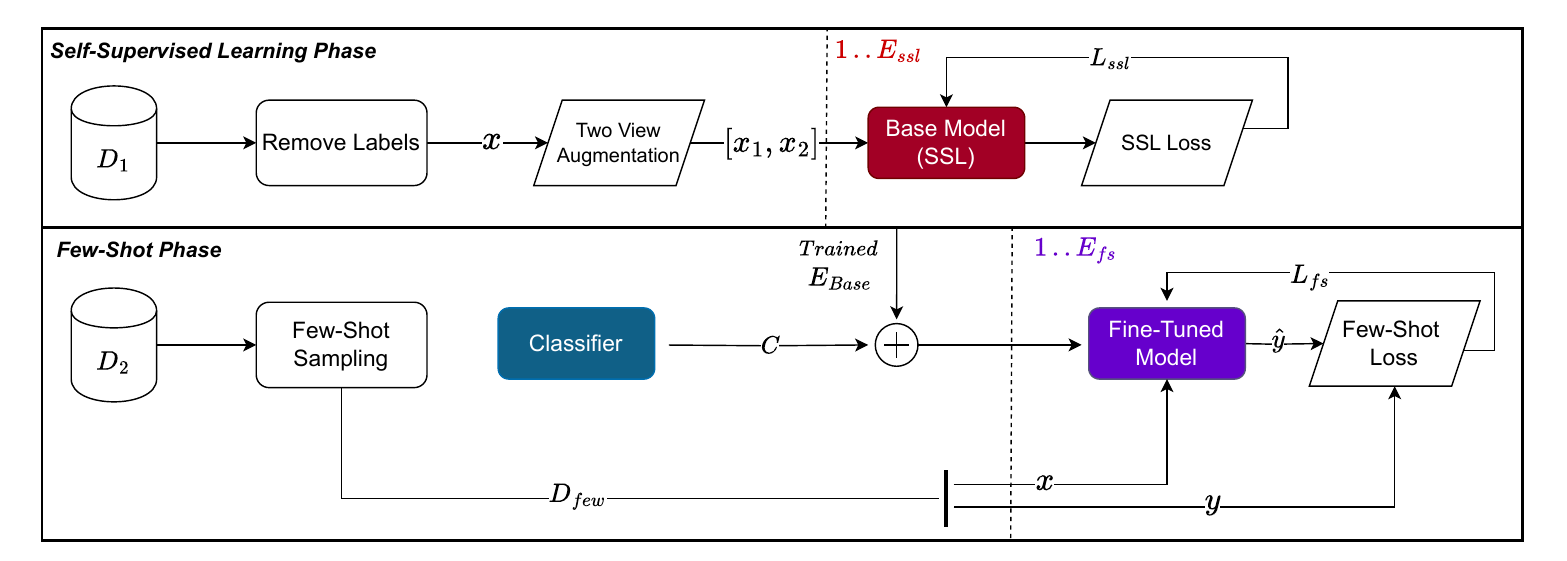}
    \caption{Graphical illustration of SSL followed by few-Shot learning.}
    \label{fig:SSL_SSL_Linear}
\end{figure*}

\section{Experiments and Discussions}
\label{sec:experiments}

This section details the experimental setup for evaluating SSL algorithms and fully supervised baselines using three Wi-Fi sensing datasets: WiMANS, UT-HAR, and SignFi. {The choice of these datasets reflects both historical relevance and the need to validate our methods across diverse settings. UT-HAR and SignFi are among the earliest and most widely used datasets for evaluating HAR, offering stable benchmarks and well-defined single-user, single-label tasks. In contrast, WiMANS is a newly introduced dataset that focuses on multi-user and multi-activity scenarios, enabling a more challenging multi-label classification task.}

{To evaluate the effectiveness of SSL under constrained label availability, we conduct 5-shot and 10-shot fine-tuning experiments on all three datasets. In addition, we explore the generalizability of learned representations through transfer learning between SignFi and WiMANS. The rationale behind this choice stems from the fact that SignFi, due to its large size and fine-grained gestures, serves as a strong candidate for SSL pretraining. WiMANS, being more complex and recently introduced, is well-suited as a downstream target to test whether SSL pretraining on an older, large-scale dataset can enable comparable performance in a more challenging setting with minimal supervision.}

Fully supervised models, trained on the complete datasets, serve as an upper bound for comparison to gauge the efficacy of self-supervised approaches in low-label regimes.

\subsection{Algorithm: SSL with Few-Shot Fine-Tuning}

We integrated SSL with few-shot fine-tuning to make the most of limited labelled data. For SSL training, a primary dataset $\mathcal{D}_1$ is used, while a second dataset $\mathcal{D}_2$ is reserved for few-shot fine-tuning in transfer learning scenarios. Key parameters include the SSL learning rate $\alpha_{ssl}$, the few-shot learning rate $\alpha_{fs}$, the number of epochs $E_{ssl}$ and $E_{fs}$, batch size $B$, and the number of labelled samples per class $n$. The framework supports various SSL methods, such as SimCLR, SimSiam, VICReg, and Barlow Twins, with a train-test split ratio $r$.
In supervised learning, the model is trained directly on labelled data from $\mathcal{D}_1$ using single-view transformations. For SSL, the model learns representations by generating two-views of each input sample. These augmented views are processed through the model using a loss function specific to the chosen SSL method, with labels ignored. After SSL training, if few-shot fine-tuning is required ($n > 0$), only the encoder $E_base$ passed while freezing it is weights, and a new classifier is trained using $n$ labelled samples per class from the training subset of $\mathcal{D}_2$, denoted as $\mathcal{D}_{2_{train}}$. This process facilitates effective transfer learning, allowing representations learned from $\mathcal{D}_1$ to be adapted to a new task using minimal labelled data.

\begin{algorithm}[h!]
\caption{SSL with Few-Shot Fine-Tuning}
\label{alg:ssl_fewshot}
\begin{algorithmic}[1]
\STATE \textbf{Input:} $\mathcal{D}_1$, $\mathcal{D}_2$ (datasets), $\alpha_{ssl}$, $\alpha_{fs}$ (learning rates), 
\STATE \hspace{1cm}$E_{ssl}$, $E_{fs}$ (epochs), $n$ (shots per class), $B$ (batch size), $alg \in \{\text{SimCLR}, ...\}$, $r$ (split ratio)
\STATE \textbf{Data Preparation:}
\STATE $\mathcal{D}_{1_{train}}, \mathcal{D}_{1_{test}} \leftarrow \text{Split}(\mathcal{D}_1, r)$
\STATE $\mathcal{D}_{2_{train}}, \mathcal{D}_{2_{test}} \leftarrow \text{Split}(\mathcal{D}_2, r)$
\STATE \textbf{Part 1: Direct Supervised Learning}
\STATE Initialize model $\mathbf{M}$
\FOR{$e = 1$ \textbf{to} $E_{ssl}$}
    \FORALL{$(\mathbf{x}, \mathbf{y}) \in \text{DataLoader}(\mathcal{D}_{1_{train}}, B)$}
        \STATE $\mathbf{x}_{aug} \leftarrow \text{Transform}(\mathbf{x})$
        \STATE $\mathcal{L}_{sup} \leftarrow \text{Loss}(\mathbf{M}(\mathbf{x}_{aug}), \mathbf{y})$
        \STATE Update $\mathbf{M}$ using $\alpha_{ssl}$ and $\mathcal{L}_{sup}$
    \ENDFOR
\ENDFOR
\STATE \textbf{Part 2: SSL}
\STATE Initialize encoder $\mathbf{E}$ for chosen $alg$
\FOR{$e = 1$ \textbf{to} $E_{ssl}$}
    \FORALL{$\mathbf{x} \in \text{DataLoader}(\mathcal{D}_{1_{train}}, B)$}
        \STATE $\mathbf{x}_1, \mathbf{x}_2 \leftarrow \text{TwoCropTransform}(\mathbf{x})$
        \STATE $\mathcal{L}_{ssl} \leftarrow \text{SSLLoss}_{alg}(\mathbf{E}, \mathbf{x}_1, \mathbf{x}_2)$
        \STATE Update $\mathbf{E}$ using $\alpha_{ssl}$ and $\mathcal{L}_{ssl}$
    \ENDFOR
\ENDFOR
\STATE \textbf{Part 3: Few-Shots Fine-Tuning}
\IF{$n > 0$}
    \STATE $\mathbf{E}_{base} \leftarrow \text{RemoveProjectionHead}(\mathbf{E})$
    \STATE Freeze $\mathbf{E}_{base}$
    \STATE Initialize classifier $\mathbf{C}$
    \STATE $\mathcal{D}_{few} \leftarrow \text{SampleNPerClass}(\mathcal{D}_{2_{train}}, n)$
    
    \FOR{$e = 1$ \textbf{to} $E_{fs}$}
        \FORALL{$(\mathbf{x}, \mathbf{y}) \in \text{DataLoader}(\mathcal{D}_{few}, B)$}
            \STATE $\mathbf{z} \leftarrow \mathbf{E}_{base}(\mathbf{x})$
            \STATE $\mathcal{L}_{fs} \leftarrow \text{Loss}(\mathbf{C}(\mathbf{z}), \mathbf{y})$
            \STATE Update $\mathbf{C}$ using $\alpha_{fs}$ and $\mathcal{L}_{fs}$
        \ENDFOR
    \ENDFOR
    \RETURN Evaluate($\mathbf{E}_{base}, \mathbf{C}, \mathcal{D}_{2_{test}}$)
\ENDIF
\RETURN $\mathbf{E}$
\end{algorithmic}
\end{algorithm}

\subsection{Data Augmentation Techniques}

The TwoCropTransform function in \textbf{Algorithm ~\ref{alg:ssl_fewshot} (line 16)} implements two augmentations. For input $\mathbf{x} \in \mathbb{R}^{H \times W \times C}$, Gaussian noise augmentation $\mathcal{T}_g$ is defined as:
\begin{equation}
    \mathcal{T}_g(\mathbf{x}) = \mathbf{x} + \epsilon, \text{ where } \epsilon \sim \mathcal{N}(0, 0.01^2)
\end{equation}
Random masking $\mathcal{T}_m$ sets randomly selected 10\% of input values to zero:
\begin{equation}
    \mathcal{T}_m(\mathbf{x})_i = \begin{cases} 
    0 & \text{with probability } 0.1 \\
    \mathbf{x}_i & \text{with probability } 0.9
    \end{cases}
\end{equation}

\begin{table}[ht!]
    \centering
    \caption{Training Parameters for SSL and Supervised Learning}
    \small
    \label{tab:parameters}
    \begin{tabularx}{0.48\textwidth}{X c c}
        \toprule
        \textbf{Parameters} & \textbf{SSL algorithmss\textsuperscript{*}} & \textbf{Supervised Baseline} \\ 
        \midrule
        Training Epochs & 50 & 50 \\ 
        Learning Rate & 0.01 & 0.001 \\
        Batch Size & 64 & 64 \\
        Few-shot Evaluation Epochs & 20 & -- \\
        Fine-tuning Learning Rate & 0.001 & -- \\
        \bottomrule
    \end{tabularx}
    \begin{center}
    \small{\textsuperscript{*}SSL algorithmss: Barlow Twins, SimCLR, SimSiam, VICReg}
    \end{center}
\end{table}

\subsection{Experimental Setup}

\subsubsection{SSL and Supervised Training}

We evaluate four SSL algorithms following \textbf{Algorithm~\ref{alg:ssl_fewshot}} using each dataset independently as $\mathcal{D}_1$ and $\mathcal{D}_2$, since the experiments are not transfer learning scenarios. The datasets used include WiMANS (6 users, 9 activities), UT-HAR (7 activities), and SignFi (sign language with 276 classes). Each experiment involves fine-tuning with $n=5$ and $n=10$ shots. Parameter updates are performed using the Adam optimizer with weight decay $\lambda=10^{-4}$, as follows:

\begin{equation} \theta_{t+1} = \theta_t - \eta \cdot \frac{\hat{m}_t}{\sqrt{\hat{v}_t} + \epsilon} - \lambda\theta_t \label{eq:adam} \end{equation}

Here, $\hat{m}_t$ and $\hat{v}_t$ represent the bias-corrected first and second-moment estimates, respectively. The training parameters specific to each dataset are provided in \textbf{Table~\ref{tab:parameters}}.

\subsubsection{Transfer Learning}
Using the same\textbf{ Algorithm~\ref{alg:ssl_fewshot}}, we test the generalization between different task domains using $\mathcal{D}_1=\text{SignFi}$ and $\mathcal{D}_2=\text{WiMANS}$. We employ SimCLR, SimSiam, BarlowTwins, and VICReg as our SSL algorithms with frozen encoder $\mathbf{E}_{backbone}$ (line 22) and data augmentation strategies. The training parameters are outlined in \textbf{Table~\ref{tab:transfer}}.

\begin{table}[ht!]
    \centering
    \caption{Training Parameters for Transfer Learning from SignFi to WiMANS}
    \small
    \label{tab:transfer}
    \begin{tabular}{l c}
        \toprule
        \textbf{Parameters} & \textbf{Values} \\ 
        \midrule
        SSL Training Epochs ($E_{ssl}$) & 50 \\ 
        SSL Learning Rate ($\alpha_{ssl}$) & 0.01 \\
        SSL Batch Size ($B$) & 128 \\
        Few-shot Evaluation Epochs ($E_{fs}$) & 100 \\
        Few-shot Learning Rate ($\alpha_{fs}$) & 0.001 \\
        Few-shot Batch Size ($B$) & 64 \\
        Number of Shots ($n$) & 5, 10 \\ 
        \bottomrule
    \end{tabular}
\end{table}

\subsection{Model Architecture}

Following \textbf{Algorithm~\ref{alg:ssl_fewshot}}, our architecture consists of encoder $\mathbf{E}$, projection head $g$ (as well as optional prediction head $h$ for SimSiam), and classifier $\mathbf{C}$.
The encoder maps the input $\mathbf{x} \in \mathbb{R}^{H \times W \times C}$, to 128-dimensional embeddings $\mathbf{E}: \mathbb{R}^{H \times W \times C} \rightarrow \mathbb{R}^{128}$.
This is achieved using a convolutional neural network with three sequential layers:
\begin{align*}
    f_1 & :\quad \mathbb{R}^{H \times W \times 1} \xrightarrow{\hspace{0.5em}(k=27,\, s=7)\hspace{0.5em}} \mathbb{R}^{h_1 \times w_1 \times 32} \\
    f_2 & :\quad \mathbb{R}^{h_1 \times w_1 \times 32} \xrightarrow{\hspace{0.5em}(k=15,\, s=3)\hspace{0.5em}} \mathbb{R}^{h_2 \times w_2 \times 64} \\
    f_3 & :\quad \mathbb{R}^{h_2 \times w_2 \times 64} \xrightarrow{\hspace{0.5em}(k=7,\, s=1)\hspace{0.5em}} \mathbb{R}^{h_3 \times w_3 \times 128}
\end{align*}

where $k_i, s_i$ denote kernel size and stride. The largest configuration for WiMANS. However, SignFi and UT-HAR use reduced kernels with the same channel progression. The SSL projection head maps encoder outputs as $g: \mathbb{R}^{128} \rightarrow \mathbb{R}^{128}$:
\begin{equation}
   g(\mathbf{z}) = W_2 \cdot \text{ReLU}(\text{BN}(W_1\mathbf{z}))
\end{equation}
where $W_1 \in \mathbb{R}^{512 \times 128}$, $W_2 \in \mathbb{R}^{128 \times 512}$. For downstream tasks in \textbf{(Algorithm~\ref{alg:ssl_fewshot}, lines~21--35)}, the classifier $\mathbf{C}: \mathbb{R}^{128} \rightarrow \mathbb{R}^{N_c}$ maps the frozen encoder outputs to classes $N_c$. For supervised mode \textbf{(Algorithm~\ref{alg:ssl_fewshot}, lines~4--12)}, the complete architecture, including the encoder, projection head, and classifier, is combined into one neural network.

\begin{table*}[ht]
\centering
\caption{Computational Cost Summary of SSL Methods for 1-Batch Training and Deployment Across Datasets}
\label{tab:comp_cost_summary}
\small
\begin{tabular}{llcccccc}
\toprule
\textbf{Dataset} & \textbf{Method} & \textbf{Phase} & \textbf{Time (ms)} & \textbf{Peak Mem (MB)} & \textbf{FLOPs (M)} & \textbf{Params (M) (Size MB)} \\
\midrule
\multirow{5}{*}{SignFi}
 & SimCLR         &  Training           & 9.7   & 266.2  & 122  & 0.70 (2.69) \\
 & SimSiam        &  Training           & 9.4   & 283.6  & 243  & 0.60 (2.32) \\
 & BarlowTwins    &  Training           & 9.8   & 283.9  & 244  & 0.70 (2.69) \\
 & VICReg         &  Training           & 10.4  & 283.9  & 244  & 0.70 (2.69) \\
 & Encoder\& Classifier & Inference  & 4.6   & 283.0  & 122  & 0.51 (1.96) \\

\midrule
\multirow{5}{*}{UT-HAR}
 & SimCLR         &  Training           & 4.3   & 123.4  & 80   & 0.24 (0.92) \\
 & SimSiam        &  Training           & 3.9   & 134.6  & 160  & 0.26 (0.98) \\
 & BarlowTwins    &  Training           & 4.2   & 134.5  & 160  & 0.24 (0.92) \\
 & VICReg         &  Training           & 4.9   & 134.5  & 160  & 0.24 (0.92) \\
 & Encoder \& Classifier  & Inference  & 1.9   & 125.2  & 80   & 0.18 (0.67) \\

\midrule
\multirow{5}{*}{WiMANS}
 & SimCLR         &  Training           & 115.7 & 6828.7 & 844  & 1.02 (3.89) \\
 & SimSiam        &  Training           & 115.2 & 7227.9 & 1689 & 1.04 (3.96) \\
 & BarlowTwins    &  Training           & 115.7 & 7227.8 & 1689 & 1.02 (3.89) \\
 & VICReg         &  Training           & 116.2 & 7227.8 & 1689 & 1.02 (3.89) \\
 & Encoder \& Classifier & Inference  & 57.5  & 7219.0 & 844  & 0.98 (3.74) \\

\bottomrule
\end{tabular}
\end{table*}

\subsection{Computational Analysis}

{ \textbf{Table~\ref{tab:comp_cost_summary}} reports per-batch FLOPs, peak
memory usage, training/inference time, and parameter count for each SSL method across the three
evaluated datasets. Our analysis reveals that, despite architectural similarities, SimCLR consistently requires fewer FLOPs than other methods (e.g., 122M vs. 243–244M on SignFi), due to its simpler projection design. However, all four SSL methods exhibit comparable training durations, likely due to shared encoder backbones and GPU-accelerated batch processing. We also observe that memory requirements and inference times scale notably with dataset complexity, with WiMANS requiring over 7GB peak memory, compared to just 1.2GB for UT-HAR.}

{ Despite SimCLR requiring approximately half the FLOPs of other approaches (122M vs. 243-244M for SignFi), training times remain consistent across methods (within $\pm$0.5ms per dataset), attributable to our optimized CNN architecture. Inference latency varies significantly by dataset complexity: 4.6ms for SignFi, 1.9ms for UT-HAR, and 57.5ms for WiMANS—the latter experiencing a 12-30 $\times$ performance penalty due to higher dimensionality and its 54-way classification head. Memory utilization follows similar patterns across datasets (283MB, 125MB, and 7219MB respectively). SimCLR demonstrates superior deployment characteristics with minimal computational overhead while maintaining competitive accuracy, rendering it optimal for resource-constrained sensing applications.}

\subsection{Results and Discussions}
\textbf{Table \ref{tab:accuracy_all_datasets}} presents a quantitative comparison of the accuracy of SSL methods followed by few-shot learning across the WiMANS, UT-HAR, and SignFi datasets. Additionally, \textbf{Table~\ref{tab:accuracy_all_datasets}} includes the accuracy of a fully supervised approach trained on the entire datasets for comparison.

\begin{table*}[h!]
    \centering
    \caption{Accuracy of SSL and Supervised paradigms on WiMANS, UT-HAR, and SignFi Datasets}
    \small
    \label{tab:accuracy_all_datasets}
    \newcolumntype{C}{>{\centering\arraybackslash}X}
    \begin{tabularx}{\linewidth}{l C C C C C C C}
        \toprule
        \textbf{Model} & \textbf{WiMANS (5-shot)} & \textbf{WiMANS (10-shot)} & \textbf{UT-HAR (5-shot)} & \textbf{UT-HAR (10-shot)} & \textbf{SignFi (5-shot)} & \textbf{SignFi (10-shot)} & \textbf{Full Dataset Acc. \%} \\ 
        \midrule
        SimCLR        & 53.31 & 56.64 & 40.6 & 41.4 & 94.96 & 95.47 & - \\ 
        VICReg        & 53.44 & 56.40 & 33.76 & 34.44 & 74.13 & 78.80 & - \\ 
        Barlow Twins  & 51.48 & 55.43 & 36.72 & 38.52 & 91.96 & 92.81 & - \\ 
        SimSiam       & 46.38 & 55.98 & 25.0 & 23.44 & 20.56 & 23.71 & - \\ 
        Supervised (UT-HAR) & - & - & - & - & - & - & 99.40 \\ 
        Supervised (WiMANS) & - & - & - & - & - & - & 56.47 \\ 
        Supervised (SignFi) & - & - & - & - & - & - & 95.58 \\
        \bottomrule
    \end{tabularx}
    \vskip 0.5em
    \footnotesize
        \textit{Note:} Accuracy is averaged across five distinct seeds.
\end{table*}


\subsubsection{WiMANS Results}

On the WiMANS dataset, SSL algorithms perform competitively when compared to a fully supervised approach. Specifically, SimCLR achieves an accuracy of 56.64\% in the 10-shot setting, slightly surpassing the supervised baseline of 56.47\%. This indicates that SSL methods, particularly SimCLR, are adept at capturing the complex activity patterns inherent in WiMANS data.

\subsubsection{UT-HAR Results}

For UT-HAR evaluation the highest accuracy is observed with Barlow Twins, which achieves 36.72\% in the 5-shot setting and 38.52\% in the 10-shot setting. VICReg follows with 33.76\% and 34.44\% accuracy for 5-shot and 10-shot settings, respectively. The limited size and simplicity of the UT-HAR dataset likely contribute to overfitting, hindering the effectiveness of contrastive learning approaches like SimCLR, which depend on diverse negative pairs to learn meaningful representations.

\subsubsection{SignFi Results}

The SignFi dataset presents a different challenge, utilizing raw CSI data across 276 sign language classes. After preprocessing, which involved calculating phase and amplitude information to form input tensors of shape $3 \times 200 \times 60$ (antennas × time steps × subcarriers), SSL algorithms showcased strong performance. SimCLR achieves 94.96\% and 95.47\% accuracy in the 5-shot and 10-shot settings, respectively, closely approaching the supervised baseline of 95.58\%. Barlow Twins also performs robustly, with accuracies of 91.96\% and 92.81\%. These results suggest that the rich motion patterns in sign language gestures provide ample features for SSL algorithms to learn effective representations.

All algorithms were evaluated under uniform parameters detailed in \textbf{Table \ref{tab:parameters}}, utilizing 5-shot and 10-shot linear evaluations with a frozen encoder.

\begin{table}[h!]
    \centering
    \caption{Test Accuracy on WiMANS (5-shot and 10-shot) after pretraining on SignFi}
    \label{tab:pretrain_transfer_wimans}
    \begin{tabular}{lcc}
        \toprule
        & \multicolumn{2}{c}{\textbf{SignFi Pretrained}} \\
        \cmidrule(lr){2-3}
        \textbf{SSL Method} & \textbf{5-shot (\%)} & \textbf{10-shot (\%)} \\
        \midrule
        SimCLR        & 36.72 & 39.87 \\
        VICReg        & 37.08 & 39.70 \\
        Barlow Twins  & 35.21 & 37.72 \\
        SimSiam       & 34.42 & 37.59 \\
        \bottomrule
    \end{tabular}
    \vskip 0.5em
    \footnotesize
    \textit{Note:} Accuracy is averaged across five distinct seeds.
\end{table}

\subsection{WiMANS Transfer Learning Results}
\label{sec:transfer_wimans}

To assess the efficacy of transfer learning, SSL algorithms pre-trained on SignFi were fine-tuned on WiMANS with limited labelled data. While the fully supervised WiMANS model achieves an accuracy of 56.47\%, SSL algorithms transferred from SignFi obtained up to 39.87\% accuracy with 10-shot training, capturing approximately 70.65\% of the fully supervised performance. These results highlight that transfer learning from SignFi provides a reasonable approximation in accuracy for WiMANS when labelled data is scarce, though it does not surpass supervised or SSL methods trained directly on WiMANS.
\textbf{Table~\ref{tab:pretrain_transfer_wimans}} presents the test accuracies for each SSL algorithm after SignFi pretraining. SimCLR and VICReg showed the highest accuracies, indicating their relative effectiveness in this transfer learning setup.

\subsection{Analysis and Insights}
\label{sec:analysis_insights}

Results across Wi-Fi sensing datasets highlight that the effectiveness of SSL methods is heavily influenced by dataset characteristics, including noise, structure, and class diversity.

The SignFi dataset, collected in a controlled lab environment, allowed SSL algorithms like SimCLR and Barlow Twins to achieve near-supervised performance due to its consistent and structured data. Similarly for WiMANS, despite its low baseline attributed to the complexities of multi-user scenarios. In contrast, the less diverse patterns in UT-HAR posed challenges for SSL methods, leading to moderate performance. These findings emphasize the critical role of well-structured data in enabling effective representation learning.

SimSiam struggled across datasets, particularly with SignFi's 276-class setup, due to its sensitivity to diverse classes. Suboptimal hyperparameters or insufficient regularization likely contributed to its collapse. VICReg showed modest improvements with additional shots, reflecting its sensitivity to small sample sizes and reliance on sufficient fine-tuning data for better generalization.

Transfer learning across different tasks demonstrated potential for generalization in Wi-Fi sensing. However, the use of few-shot learning often led to overfitting early in training, a limitation partly attributed to the nature of Wi-Fi datasets and current SSL methods.

{From a computational perspective, training SSL on resource-constrained devices, even though theoretically promising, is not optimal due to its need for larger datasets, longer training times, and substantial memory demands as observed with WiMANS. Our analysis suggests that richer and more complex datasets will scale computational requirements exponentially, posing significant challenges for edge deployment of SSL training processes.}

\section{Challenges and Opportunities}
\label{sec:challenges}

\subsection{Environmental Dependency}
\label{subsec:env_dep}
A critical challenge in Wi-Fi sensing lies in the environmental distribution bias that hinders the scalability and reliability of DL models. While many studies in this field report high accuracy, their evaluations often assume that training and testing data are collected under the same environmental conditions. This assumption limits the generalizability of these models when applied to novel domains, such as different locations, channel conditions, or device configurations. Domain variability, which includes differences in time, location, hardware, and protocol settings, significantly impacts model performance. As highlighted \cite{elmaghbub2024distinguishable}, the performance of DL models can degrade substantially when tested in domains with different wireless channel conditions, receiver hardware, or spatial configurations than those used during training. 
To mitigate these challenges, recent research proposed several solutions, including the generation of environment-agnostic features like the BVP representation \cite{9516988}, domain adaptation techniques \cite{8487345}, {Doppler velocity-based representations} \cite{10734771}, and advanced architectures such as matching neural networks \cite{shi2022environment}. Despite these advancements, achieving robust performance across diverse environments remains an open challenge.

{
SSL offers a promising solution to tackle the environmental distribution bias in Wi-Fi sensing. By leveraging large amounts of unlabeled CSI data, SSL methods can learn robust feature representations that capture the intrinsic patterns of human activities while being less sensitive to environmental variations such as multipath effects or interference, focusing on activity-related features rather than environment-specific noise. Recent work, such as the AutoFi framework~\cite{yang2022autofi}, demonstrates the efficacy of geometric self-supervised learning, utilizing low-quality, randomly captured CSI samples to achieve high performance in tasks like human gait and gesture recognition across diverse environments. Similarly, SSL has been shown to enhance robustness to input corruptions and out-of-distribution data, which is critical for handling environmental variability in Wi-Fi sensing~\cite{hendrycks2019using}. This ability to learn generalizable representations from unlabeled data makes SSL particularly suitable for applications where labeled data is scarce and environmental conditions vary widely.}

{
To ensure the long-term stability of SSL models in dynamic environments, several adaptation strategies can be employed. One effective approach is fine-tuning the pre-trained SSL model with a small set of labeled data from the target environment, allowing the model to adjust its learned representations to specific conditions while retaining general knowledge. For example, the FewSense system~\cite{yin2022fewsense} shows that fine-tuning a feature extractor with just a few samples from the target domain achieves accuracies of up to $96.5\%$ on datasets like Widar for activity recognition in unseen environments. Additionally, domain adaptation techniques tailored for Wi-Fi sensing, such as those surveyed in~\cite{chen2023cross}, align feature distributions between source and target domains, enhancing robustness to environmental shifts. Methods like AutoFi integrate few-shot calibration with SSL, enabling effective knowledge transfer to new tasks with minimal labeled data~\cite{yang2022autofi}. These strategies enable maintaining model performance in diverse, real-world deployment scenarios, such as smart homes or offices, where environmental conditions evolve over time.
}

\subsection{Open Access Datasets}

A notable concern within the Wi-Fi sensing literature is the lack of standardized and publicly accessible datasets for various Wi-Fi sensing tasks. To our knowledge, only the datasets mentioned in \textbf{Table \ref{datasets_table}} are available, yet they are seldom used for benchmarking in most publications. Currently, many researchers in this field opt to create their datasets using CSI extraction platforms, such as the 802.11n CSI Tool. Addressing this issue by establishing more comprehensive and publicly available datasets will facilitate the comparison and evaluation of different methods, ultimately fostering progress in the Wi-Fi sensing domain.

\subsection{Synchronization and Annotations}
{Another challenge in CSI collection, particularly in setups with multiple access points, is the issue of synchronization. Accurate synchronization is critical to ensure that the data collected from different access points is temporally aligned. Any misalignment can lead to distorted interpretations of activities and hinder the performance of data fusion techniques. The lack of standardized tools or protocols for achieving precise synchronization across multiple devices exacerbates this challenge. As a result, researchers often rely on custom solutions, which may not generalize well to other experimental setups or real-world deployments.}

{Moreover, labeling and segmenting CSI data poses a considerable challenge. Each segment of CSI data must correspond to a specific activity trial, requiring meticulous annotation, which is time-consuming and prone to errors. Unlike visual data, where activities can often be intuitively labeled using video footage, CSI data lacks a direct and interpretable representation of human activities. Ensuring accurate correspondence between the CSI signals and the performed activities makes it challenging. Additionally, errors in segmentation or labeling can significantly impact model training and evaluation, underscoring the need for robust techniques or automated tools to streamline these processes.}

\subsection{Privacy Preservation in Self-Supervised Learning}

{Although Wi-Fi sensing is often regarded as more privacy-friendly compared to vision-based methods, applying SSL introduces new privacy challenges. Unlike supervised learning, which focuses on task-specific feature extraction, SSL aims to learn broad representations that capture all distinguishing characteristics in the data. This increases the risk of inadvertently encoding sensitive user behaviors such as individual movement patterns, breathing rates, or environmental contexts. Recent work has shown that SSL models are vulnerable to various attacks, including extraction and inference attacks \cite{dziedzic2022difficulty}.}

{Furthermore, Wi-Fi signals can penetrate physical barriers and are inherently open to interception, enabling the extraction of sensitive information without explicit user consent. For instance, WiHear \cite{wang2014we} demonstrates the feasibility of inferring mouth motions and even conversations from Wi-Fi reflections. These risks emphasize that the modality itself does not fully eliminate privacy threats when powerful learning models are involved. To mitigate these concerns, differential privacy (DP) techniques have been proposed. Recent efforts, such as DP-SSLoRA \cite{yan2024dp}, integrate DP into SSL frameworks by injecting controlled noise during training, reducing the risk of sensitive information leakage while preserving model effectiveness. Nevertheless, achieving a robust trade-off between privacy guarantees and sensing performance remains a significant open research problem in the domain of Wi-Fi-based SSL.}

\subsection{Lightweight Model Design}

Among the major issues in Wi-Fi sensing applications is the need for DL models to adequately interpret time-dependent datasets. These models often demand substantial computational resources, limiting their real-world deployment—especially for latency-sensitive tasks like respiration monitoring \cite{9831898}. Traditional cloud-based inference introduces network delays and reliability constraints, which are not ideal for real-time applications.

Edge computing offers partial relief by moving computation closer to the data source; however, it still requires relatively powerful hardware. A more viable solution for highly constrained devices is the use of TinyML \cite{capogrosso2024machine}, which enables deployment of optimized ML models on microcontrollers. These models operate with minimal memory and power, enabling fast, local inference.
Lightweight CSI compression techniques, such as \cite{yang2022efficientfi}, can also reduce the bandwidth and processing burden, complementing TinyML for real-time, distributed Wi-Fi sensing. Together, these advances create a foundation for efficient edge deployment.

{As the need for SSL usage grows and showcases its potential, it still requires large computational power, making it suboptimal for training on resource-constrained devices \cite{liu2022bringing}. However, recent efforts like \cite{xiao2021toward} have demonstrated that specific SSL methods—such as self-supervised GAN-based architectures—can be adapted to the edge by generating synthetic data and using lightweight classifiers for distribution-aware learning.}
{Other approaches, such as \cite{saeed2021sense,azghan2025cudle}, utilize SSL as a pre-training stage before fine-tuning smaller models on the edge. These frameworks show significant label efficiency and improved generalization in low-data regimes. However, current solutions still fall short of addressing real-time edge training. Possible directions include leveraging TinyML for on-device inference, split learning to offload partial training to the cloud, and distributed learning for collaborative low-power devices.}

\subsection{Limitations in Existing Datasets}

{Wi-Fi sensing datasets, particularly those utilizing CSI for human activity recognition, face numerous limitations that hinder their generalizability and deployment in real-world applications \cite{8067693, 8866726, 8367378, huang2024wimans, 10.1145/3161183}. A significant concern is their limited generalization capacity, often rooted in the small number of participants involved. Most datasets are collected from homogeneous groups in constrained environments, failing to capture demographic and behavioral diversity necessary for robust models \cite{Li_Cui_Wang_Zhang_Chen_Wu_2021, 9516988, 10177905, ZHURAVCHAK202259}. This lack of diversity restricts the applicability of these models across domains, impeding efforts like domain adaptation.}

{Additionally, the datasets typically emphasize a narrow set of activities to reduce data collection time, which restricts their ability to model spontaneous, complex behaviors \cite{8067693, 8866726, 8367378, huang2024wimans, 10.1145/3161183}. Although some datasets, such as WiMANS, address multi-user recognition, performance remains low due to the challenges inherent in modeling multi-user interference and interaction \cite{huang2024wimans}.}

{Another recurring issue is that most datasets are collected exclusively in controlled indoor environments, leading to a domain bias that hinders model generalization to outdoor or dynamic real-world settings \cite{10.1145/3191755, 9235578, 10144501}. Moreover, these datasets are highly sensitive to environmental noise and changes in layout or reflectors, necessitating significant preprocessing to extract stable features \cite{8487345, 8970452}. This dependency on environment-specific conditions raises concerns regarding the robustness and scalability of models trained on them.}

{Cross-device generalization is another underexplored yet critical limitation. Models trained on data from a specific hardware setup often fail to generalize to different chipsets, antenna configurations, or vendors due to hardware-induced variations in CSI measurements \cite{8067693, 8367378}. Despite the practical significance of this challenge, very few datasets offer cross-device recordings to facilitate studying hardware-agnostic feature representations.}

{Finally, most datasets assume line-of-sight (LOS) conditions, whereas real environments often involve non-line-of-sight (NLOS) scenarios. The performance of CSI-based recognition typically degrades under NLOS settings due to increased multipath effects and signal attenuation, which remain insufficiently addressed in existing collections \cite{10.1145/3161183, s20144025, 8970452}.}

{In summary, while existing CSI datasets have enabled substantial progress, their limitations in diversity, environmental robustness, hardware heterogeneity, and multi-user capability necessitate the development of more comprehensive, standardized, and realistic datasets to ensure reliable Wi-Fi sensing across practical deployment scenarios}

\subsection{Multi-modal Learning}

{Multi-modal learning enhances decision-making by combining diverse modalities. For example, WiVi \cite{Zou_2019_CVPR_Workshops} shows the benefit of integrating Wi-Fi sensing with visual recognition for HAR, combining the strengths of both streams. Similarly, \cite{guo2024human} demonstrates that fusing CSI and IMU signals achieves higher accuracy and robustness compared to unimodal setups. By extracting joint features and leveraging complementary information, multi-modal learning improves model reliability in complex environments. However, its application in real-time or edge settings remains limited due to increased computational demands. As discussed in \cite{10630605}, current fusion strategies often introduce complex architectures, making them challenging for latency- and energy-constrained systems. Additionally, these models typically assume information overlap between modalities, synchronized extraction of different modalities, and require a similar quantity of samples across modalities that may not always exist in practice.} { Moreover, in SSL settings, where training is already computationally intensive due to large unlabeled datasets, introducing additional modalities further increases complexity.  }

{Furthermore, variations in hardware and environmental conditions across modalities can degrade data quality across modalities, as noted in Section~\ref{subsec:env_dep}. Combining CSI with vision-based data, in particular, undermines key advantages of Wi-Fi sensing, namely privacy and computational efficiency \cite{zhou2023towards}. Future research should explore alternative complementary modalities beyond vision, aiming to achieve the benefits of multi-modal fusion while maintaining low computational cost and strong privacy protections.
Clearly, while multi-modal fusion can significantly enhance HAR robustness and accuracy, it also introduces a trade-off between performance and deployment feasibility. Overcoming these challenges requires the development of lightweight fusion methods and efficient SSL strategies that preserve Wi-Fi sensing’s inherent benefits.}


\section{Conclusion and Outlook}
\label{sec:conclusion}

In this survey-cum-tutorial article, we outlined various Wi-Fi standards, discussed fundamentals of CSI extraction for stationary and
mobile scenarios, and described CSI measurement tools and techniques.
We also presented a comprehensive review of the existing CSI datasets and their specifications, while showcased a simplified testbed for CSI extraction. A variety of preprocessing techniques that can enhance the performance of DL algorithms have been identified and discussed. We then provided a qualitative comparison of the existing DL methods, including supervised, unsupervised, and SSL, for CSI-based Wi-Fi sensing. Experiments are then presented to show a quantitative comparison of various datasets and SSL models.  Finally, we discussed open challenges and  opportunities.

{Wi-Fi sensing remains an active area of research, offering significant potential across diverse applications. For instance, in cross-domain fields such as healthcare and autonomous vehicles, it can enhance obstacle detection and provide feedback in scenarios where traditional cameras fall short. Advancing the hardware used for dataset collection is essential to address limitations in multi-user environments. Furthermore, key challenges like improving data quality, utilizing multimodality, and enhancing generalization across domains must be tackled.} In summary, while SSL methods are promising for Wi-Fi sensing, their success depends on dataset quality, diversity, and task complexity. Future efforts should focus on refining training strategies, addressing overfitting in transfer learning, and tailoring hyperparameters to dataset-specific challenges. { Additionally, developing mathematical models to establish a direct link between CSI dynamics and SSL representation learning would provide theoretical foundations for understanding how wireless channel variations translate into meaningful representations.}


\bibliographystyle{IEEEtran}
\bibliography{ref}


\end{document}